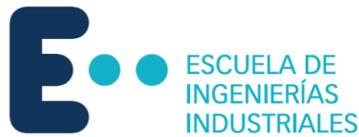
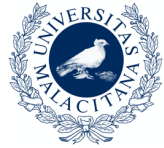

Escuela de ingenierías industriales

Ingeniería de sistemas y automática

Visión por computador

# TRABAJO FIN DE GRADO

**Desarrollo del software de interfaz de usuario para el espectrógrafo astronómico COLORES instalado en el observatorio de la Mayora**

**Grado en Ing. Electrónica, Robótica y Mecatrónica**

**Autor**: Álvaro Montoro Martínez

**Tutor**: Carlos Jesús Pérez del Pulgar Mancebo
**Cotutor**: Ignacio Pérez García

# Declaración firmada sobre originalidad del trabajo

D. Álvaro Montoro Martínez, con DNI 15514384X, estudiante del grado de Ingeniería Electrónica, Robótica y Mecatrónica cuyo Trabajo Fin de Grado *Desarrollo del software de interfaz de usuario para el espectrógrafo astronómico COLORES instalado en el observatorio de la Mayora*, **declaro bajo mi responsabilidad** ser autor del texto entregado y que no ha sido presentado con anterioridad, ni total ni parcialmente, para superar materias previamente cursadas en ésta u otras titulaciones de la Universidad de Málaga o cualquier otra institución de educación superior o de otro fin.

Así mismo, declaro no haber trasgredido ninguna norma universitaria con respecto al plagio ni a las leyes establecidas que protegen la propiedad intelectual, así como que las fuentes utilizadas han sido citadas adecuadamente.

Málaga, a 9 de Mayo de 2022

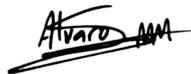

Fdo.: Álvaro Montoro Martínez
Autor del TFG

*A todas las personas que de alguna u otra manera me han ayudado a ser quien soy hoy.*

*A.M*

# Resumen

La espectroscopia se ha vuelto con los años una herramienta fundamental para los astrónomos, pues otorga información muy valiosa sobre cuerpos que están a una gran distancia de nosotros. No solo ayuda a la hora de detectar la composición química de los astros observados, sino también su distancia con respecto a nosotros, su temperatura y densidad, amen de otras muchas propiedades.

Este proyecto de fin de grado nace por tanto, de la necesidad de poner en funcionamiento el espectrógrafo (COLORES) de la estación (BOOTES 2), ubicada en la Mayora y perteneciente a la red de telescopios Burst Observer and Optical Transient Exploring System (BOOTES). Un telescopio robótico como el ubicado en BOOTES 2 posee entre sus muchas virtudes, la capacidad de realizar multitud de observaciones con un tiempo de reacción muy bajo. Esto permite, obtener una gran cantidad de datos sobre el posicionamiento y la caracterización de los cuerpos astronómicos. Con esta herramienta en funcionamiento, ahora se podrán extraer multitud de nuevos parámetros de las observaciones, dotando a esta estación de un instrumento más completo y versátil con el que obtener información de mayor interés científico.

Para esta tarea, se realizarán una serie de scripts[1]. En concreto dos, uno para la calibración del espectrógrafo y otro encargado del procesamiento de la imagen y la extracción de su espectro. Esto se llevará a cabo, mediante el software Spyder (Python), en el que además se realizarán numerosas pruebas para comprobar que el software funciona perfectamente. Una vez realizados estas pruebas, se implementará en la página Web del telescopio para su uso.

Para este propósito se utilizarán varias librerías, entre ellas Astropy que incluye un completo paquete para el manejo de datos astronómicos en Python y Matplotlib que permite la utilización de gráficos generados a partir de datos contenidos en arrays. Además, se utilizarán diversas técnicas de adquisición de imágenes como: filtrado, ajuste gaussiano, utilización de regiones de interés. Con todo ello, se optimizarán los datos extraídos del telescopio para conseguir los resultados deseados.

---

[1]Instrucciones escritas en código que sirven para ejecutar diversas funciones dentro de un programa.

# Índice general

# 1 Introducción

## 1.1 Motivación

El desarrollo de la robótica en la actualidad y la implementación de esta en las distintas áreas del conocimiento ha permitido la automatización de procesos que antes requerían de una gran cantidad de tiempo y esfuerzo, suponiendo un impedimento en el avance de la ciencia. En el ámbito de la astrofísica más concretamente, la espectroscopia se ha convertido en una herramienta indispensable para el estudio de las características y la composición de los cuerpos celestes que con cada vez más frecuencia son hallados en el universo. Operar con un espectrógrafo es un proceso en el que la precisa colocación del telescopio es clave para obtener un espectro lo más fiel posible.

Este espectro requiere de un tratamiento para conseguir datos de interés científico, es por esto que se necesita un software que se encargue de extraer la información contenida en la imagen captada por el telescopio.

## 1.2 Objetivos

Este Trabajo Final de Grado (TFG) ha sido planteado con el propósito de automatizar la toma de espectros e implementar los datos obtenidos del Compact Low Resolution Spectrograp (COLORES) como parte de la de la red de telescopios robóticos BOOTES perteneciente al Instituto de Astrofísica de Andalucía (IAA).

A partir de la recolección de espectros obtenidos operando con precisión el telescopio, estos se procesarán mediante un software desarrollado en Python e implementado en el propio servidor del telescopio para obtener una gráfica en 2 dimensiones que muestre las distintas líneas de emisión y absorción del cuerpo, además de la reconstrucción del continuo. Esto se produce cuando las interacciones de un gran número de partículas extienden todas las líneas de emisión discretas de un objeto, de modo que ya no pueden distinguirse.

## 1.3 Metodología

Para procesar de forma ordenada la información se van a seguir una serie de hitos en la elaboración del proyecto:

1. **Red BOOTES**: Se explicará tanto la labor de la red de telescopios robóticos, así como las necesidades que busca cubrir este proyecto dentro del campo de la astronomía.

2. **Caracterización del instrumento COLORES**: Una visión detallada del instrumento principal con el que se va a operar dentro de la estación BOOTES 2.

3. **Elección del software**: Aunque durante el desarrollado del grado el principal software de programación que se ha utilizado es MatLab, Python ofrece mayor flexibilidad a la hora de realizar este proyecto puesto que consta de librerías propias para trabajar con imágenes astronómicas lo cual facilita el trabajo. Además, la mayor parte del software contenido en el servidor del telescopio se ha desarrollado en este entorno de programación. Por tanto, se evitarían posibles problemas de compatibilidad.

4. **Astropy y como trabajar con .fits**: Recopilación de las librerías, funciones y métodos utilizados para la realización del proyecto.

5. **Puesta en marcha y testeo del software**: Finalmente, se comprobará si los resultados del software son correctos mediante imágenes capturadas con el telescopio de la estación BOOTES 2.

## 1.4 Estructura

1. **Introducción**: En este apartado se expone una descripción del por qué es necesario este trabajo y el objetivo que se pretende conseguir con él.

2. **Antecedentes**: Una breve explicación de la red BOOTES y de los objetivos con los que se planteó el proyecto COLORES.

3. **Descripción del instrumento**: Como su propio nombre indica, se detallarán las distintas partes del instrumental que se utiliza para la adquisición de imágenes.

4. **Adquisición de imágenes con BOOTES 2**: En este capítulo se explica el proceso para adquirir las imágenes que se van a estudiar del telescopio.

5. **Programa desarrollado**: Descripción detallada y pormenorizada del código que se ha desarrollado en Python para la obtención de los resultados deseados.

6. **Pruebas y resultados**: Puesta en marcha del software desarrollado, probándolo con distintas imágenes capturadas por el propio telescopio y resultados obtenidos.

7. **Conclusiones**: Finalmente, se comentan las conclusiones obtenidas tras la realización del trabajo y se detallan los caminos que se abren tras la elaboración de este TFG.

# 2 Antecedentes

## 2.1 Espectroscopia en la observación astronómica

Se puede definir la espectroscopia, como el estudio de la interacción entre la radiación electromagnética y la materia. En el caso de la astronomía, el objeto de estudio son las estrellas y demás cuerpos celestes y la radiación electromagnética que emiten y que llega hasta nosotros. [1]

Como se ha mencionado antes, la espectroscopia es una de las herramientas favoritas de los astrónomos para entender el universo. Cuando se habla de estrellas, galaxias y demás cuerpos estelares hay que entender que las distancias son demasiado grandes como para estudiar sus propiedades en un laboratorio. Lejos de ser un obstáculo insalvable, la espectroscopia permite gracias a la luz que se recibe de estos cuerpos distantes hallar algunas de sus propiedades. Sin embargo, para poder extraer toda la información contenida en ella se necesita dividir en sus distintas longitudes de onda (colores).

Fue Sir. Isaac Newton quien originariamente utilizó un prisma para descomponer la luz del sol, llamando al arco-iris resultante espectro, que en latín significa apariencia. No fue hasta principios del siglo XIX, cuando el astrónomo alemán Joseph von Fraunhofer y el físico del mismo país Gustav Robert Kirchhoff describieron el patrón de bandas oscuras que se observaban en el espectro continuo solar como lineas de absorción.[1] Lo interesante de este fenómeno, es que estas lineas actúan como una especie de huellas dactilares, pues imprimen en el espectro la interacción de la luz con los diferentes elementos químicos. Cada molécula genera una firma única en el espectro, que la diferencia de las demás. Descifrando estas marcas, se puede revelar información importante de cualquier cuerpo que interaccione con la luz emitiéndola o absorbiéndola.

Para clarificar más este fenómeno se va a usar un ejemplo práctico. Una estrella emite luz en todo el espectro, es lo que se conoce como continuo. Cuando la luz, blanca[2], atraviesa un prisma, forma un arco iris. Sin embargo, este es un caso ideal. Cuando la luz de una estrella atraviesa el gas de una nebulosa o sin ir tan lejos el de su propia atmósfera, determinadas longitudes de onda son absorbidas por los elementos conte-nidos en el gas, produciendo líneas oscuras sobre el continuo. Esto se conoce como un espectro de absorción, pero a su vez la energía absorbida por el gas se emite en todas direcciones, en las longitudes de onda de los elementos presentes en el gas, produciendo líneas brillantes. Esto se conoce como un espectro de emisión. Unos años tras el descubrimiento de Fraunhofer y Kirchhoff, el padre Angelo Secchi un astrónomo italiano, realizó la primera clasificación estelar según el tipo espectral. Siendo el primero en poner de manifiesto la relación entre el color de las estrellas y las líneas espectrales.

---

[1]En un espectro continuo y uniforme se llama línea de absorción a la carencia de fotones en un rango de frecuencias comparando estas con las frecuencias cercanas.

[2]Se llama así por contener la totalidad del espectro visible, entre los 400 y 700 nm, del espectro electromagnético, en el cual se mezclan todos los colores. Por eso, se ve blanco.

Hay que remarcar que los espectrógrafos actuales son mucho más sofisticados que un prisma. Los espectros obtenidos muestran la luz de forma más dispersa que en un prisa y además, la luz se registra en un Dispositivo de carga acoplada (CCD) cuyo funcionamiento se explicará en capítulos posteriores. La espectroscopia astronómica no solo es útil en cuanto al estudio de las estrellas se refieres, también se utiliza en la observación y el estudio de otros cuerpos celestes, como las nebulosas, las galaxias, los cuásares, los planetas y los asteroides o los cometas. Algunos de las características de un cuerpo que se pueden estudiar mediante su espectro son: su composición química, su temperatura interna, su densidad, información sobre el campo magnético presente en el objeto y su movimiento respecto a la tierra. Para entender el movimiento de un astro mediante la espectroscopia, se utiliza el famoso efecto Doppler.[3]

La espectroscopia ha resultado vital, sin ir más lejos, en el estudio de las galaxias. En la década de 1920 el astrónomo estadounidense Edwin Hubble observó que, aparte de las galaxias más cercanas (conocidas como el Grupo Local), todas las galaxias se alejan de la Tierra. No solo eso, sino que también comprobó que cuanto mayor era la distancia entre esa galaxia y el observador, en este caso la tierra, más rápido se alejaba. Este descubrimiento es de extremada importancia en el campo de la física, puesto que fue el primer y principal punto de apoyo de la Teoría del Big Bang.[4]

En la actualidad, hay que mencionar el desplazamiento hacia el rojo (en inglés: redshift). Explicado de una manera sencilla, ocurre cuando la radiación electromagnética, más concretamente, la luz visible que se emite o refleja desde un cuerpo, es desplazada hacia el rojo al final del espectro. Esto ocurre cuando la fuente de luz se aleja del observador de manera similar a como ocurre con el efecto Doppler. En caso contrario el desplazamiento hacia el azul significa que el cuerpo se acerca al observador. Lo interesante de este fenómeno, es que la espectroscopia astronómica actual utiliza el corrimiento al rojo para determinar el movimiento de cuerpos lejanos. Este se puede producir por diversos motivos, como la expansión métrica del espacio predicha por Hubble, o por el conocido efecto Einstein, también llamado corrimiento al rojo gravitacional.[5] En el informe Calculating Redshift via Astronomical Spectroscopy [2] se puede ver un ejemplo de ello.

---

[3] Cambio de frecuencia aparente de una onda producido por el movimiento relativo de la fuente respecto a su observador.

[4] Teoría por la cual hace 13.800 millones de años se origino el universo, a partir de un punto inicial en el que se formó la materia, el espacio y el tiempo.

## 2.2 Red BOOTES

La red BOOTES está formada por 5 observatorios astronómicos situados en puntos estratégicos de forma que su campo de visión cubre toda la bóveda celeste (España, Nueva Zelanda, China y México)[3].

Sus objetivos científicos incluyen:

1. Detección de destellos ópticos de origen cósmico.

2. Control de todo el cielo con las cámaras CASANDRA por debajo de la 10ª magnitud cada 60 segundos.

3. Control de diferentes tipos de objetos variables por debajo de la 20ª magnitud para buscar variabilidad óptica.

4. Descubrimiento de cometas, meteoros, asteroides, estrellas variables, novas y supernovas.

Para la realización de este TFG se ha utilizado el telescopio de la estación BOOTES 2 ubicado en la Estación Experimental de La Mayora, Málaga. Esto se debe a que es el único donde se encuentra ubicado el espectrógrafo COLORES con el que se va a trabajar.

Este telescopio al igual que todos los de la red tiene como punto a favor que puede ser controlado de forma remota mediante esta interfaz de usuario (Figura 2.1).

El objetivo principal de la red es observar rápidamente eventos transitorios. Se entiende por evento transitorio, cualquier proceso astronómico cuya duración esté limitada en el tiempo. Esta limitación en el tiempo suele estar ligada a la longitud de onda de la radiación, desde milisegundos en la zona más energética del espectro electromagnético hasta meses y años para ondas de radio.

---

[5]Dilatación del tiempo que ocurre cerca de objetos masivos, como agujeros negros o estrellas supermasivas.

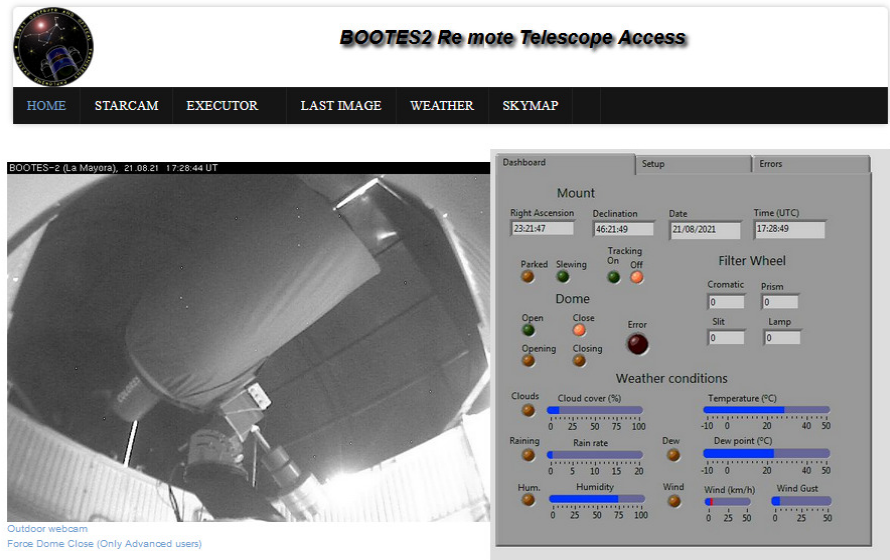

**Figura 2.1:** Visual del control del telescopio BOOTES 2

## 2.3 Objetivos del proyecto COLORES

Como se ha mencionado antes, los telescopios robóticos traen consigo unas innegables virtudes como la rapidez, la cantidad de datos que pueden recoger en poco tiempo o el abaratamiento de costes al no necesitar un operador especializado. Por otro lado, son más complejos en cuanto a su diseño y requieren de una fiabilidad máxima.

Un Faint Object Spectrograph and Camera (FOSC) tiene una única trayectoria óptica fija con un conjunto de ruedas en varias posiciones que permiten introducir o retirar una serie de rendijas, filtros y rejillas en la trayectoria de la luz. La ventaja adicional es que los elementos ópticos activos se introducen en el espacio de colimación, por lo que no se necesitan cambios de enfoque.

El objetivo con el que se diseñó COLORES era el de conseguir un instrumento que funcione bien en ambos modos (imagen y espectroscopia), que evite las partes móviles (aparte de las ruedas) y, sobre todo, que garantice que el objeto centrado en el modo de imagen sea el objeto cuyo espectro se captura en el modo de espectroscopia.

Una aplicación inmediata de este instrumento en el caso de la red BOOTES es la rápida obtención de imágenes, su localización y, eventualmente, una rápida espectroscopia de un fenómeno transitorio relacionado con un Gamma-Ray Burst (GRB). Los GRB son uno de los fenómenos más energéticos del Universo y conseguir observarlos requiere de un considerable esfuerzo científico, puesto que su duración va de los milisegundos a los minutos.

Esto crea una oportunidad única para los instrumentos de reacción rápida. Un FOSC puede hacer esto, y luego capturar un espectro con el mismo instrumento. Además del valor científico, esta información es vital para decidir si hay que activar telescopios gigantes para realizar estudios más detallados.

# 3 Descripción del instrumento

En este capítulo se explicarán las características del instrumento que se utilizará para la adquisición de imágenes en las que se basa el proyecto, así como el proceso de calibración de las mismas.

## 3.1 COLORES

COLORES es un espectrógrafo compacto y ligero de 13kg diseñado para telescopios robóticos. Se trata de un instrumento multimodo que permite al observador cambiar sin problemas entre los modos de espectroscopia de baja dispersión y de imagen directa durante una observación.

En modo espectroscópico, el instrumento dispone de cuatro largas rendijas disponibles (9.3 min long arco), de diferentes aperturas (1.1 pulg, 2.1 pulg, 3.2 pulg y 4.3 pulg) para adaptar el instrumento a la variación de la visión atmosférica.

El elemento utilizado para la dispersión de la luz es un prisma de rejilla. Se utiliza para reducir los problemas que produce en la óptica global el no tener partes articuladas (cámara o dispersor). Además, permite una configuración de visión directa, con una trayectoria óptica lineal a través del espectrógrafo.

### 3.1.1 Parámetros básicos de diseño

Dado que el colimador refractivo y la cámara son idénticos la escala del plato de COLORES sería la misma en cualquier telescopio. Sin embargo, teniendo en cuenta que la relación focal del espectrógrafo es de f/8 puede ser necesario un reductor focal[1] para los telescopios con relación focal <8 para no perder luz, y obviamente en este caso cambiar la escala del plato.

El instrumento se muestra en la (Figura 3.1), donde se detallan los principales elementos, desde la entrada hasta el detector. Los requisitos técnicos del mismo se resumen en la (Figura 3.2) [4].

---

[1]Reduce el tiempo de exposición a la mitad para capturar el mismo brillo de un objeto.

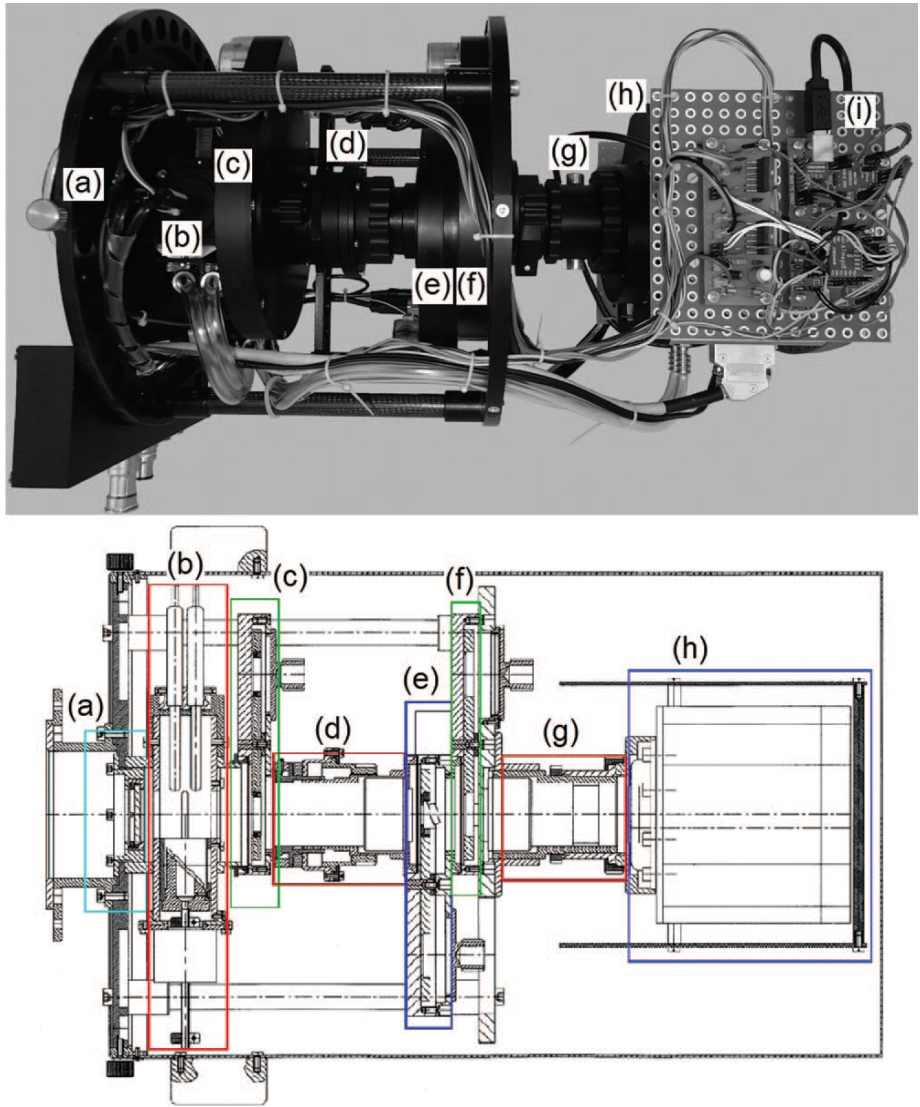

**Figura 3.1:** Fotografía y esquema detallado de COLORES. Las partes principales son: (a) entrada del instrumento, (b) sistema de calibración de longitudes de onda, (c) rueda de rendijas, (d) colimador refractivo, (e) rueda de prismas (y polarizador), (f) rueda de filtros, (g) cámara refractiva, (h) cámara CCD, y (i) sistema de control electrónico. [4]

| | |
|---|---|
| Resolución ($\lambda/\triangle\lambda$) | 100-1100 |
| Rango de longitud de onda | 0.38-1.15$\mu m$ |
| Campo de visión | 9.5 x 9.5 arc min |
| Modos de observación | Espectroscopia, imagen directa |

**Figura 3.2:** Tabla de requisitos técnicos de COLORES [4].

Por tanto, se diseñó un espectrógrafo compacto de rendija larga y baja dispersión que también funcionará como cámara para captar imágenes, un requisito esencial para realizar trabajos con GRB. El instrumento se montará en un telescopio de respuesta rápida, ubicado en una cúpula relativamente pequeña. Los principales parámetros de la configuración óptica y el rendimiento para COLORES se resumen en la (Figura 3.3).

| | |
|---|---|
| **Telescopio** | |
| Diámetro | 600 mm |
| f/relación | 8 |
| Escala | 23.27 $\mu m/arc$ seg |
| **Rendijas** | |
| Longitud | 9.3 arc min |
| Ancho | 1.1 pulg., 2.1 pulg., 3.2 pulg. y 4.3 pulg. |
| **Colimador dióptrico/cámara** | |
| Longitud focal | 50.64 mm |
| f/relación | 8 |
| Diámetro de la pupila | 6.3 mm |
| Radio de campo | 10.52º |
| **Prisma de rejilla** | |
| Prisma | CTK19, P-SF68, SF2, N-BK7 |
| Rejilla | B270 Schott |
| Ángulo de resplandor | 17.5º, 28.7º, 31.7º |
| Ranuras | 300 l/mm, 600 l/mm |
| **CCD** | |
| Píxeles activos | 1024 x 1024 |
| Tamaño de píxel | 13.3 x 13.3 $\mu m$ |

**Figura 3.3:** Características ópticas de Telescopio de Málaga (TELMA)+COLORES [4].

### 3.1.2 Lámparas de calibración

La calibración de las longitudes de onda se realiza mediante dos lámparas de calibración, una no identificada y otra de HgAr (gas argón y vapor de mercurio).

Se encuentran en el interior de COLORES y por su disposición pueden retirarse y cambiarse de forma sencilla. Como se ha visto en la (Figura 3.1), las lámparas se encuentran montadas antes de la rueda de apertura, en linea recta con la trayectoria óptica.

La ubicación de estas líneas espectrales (Figura 3.4) en el detector nos permite calcular la relación entre el píxel y la longitud de onda mediante un polinomio de orden n.

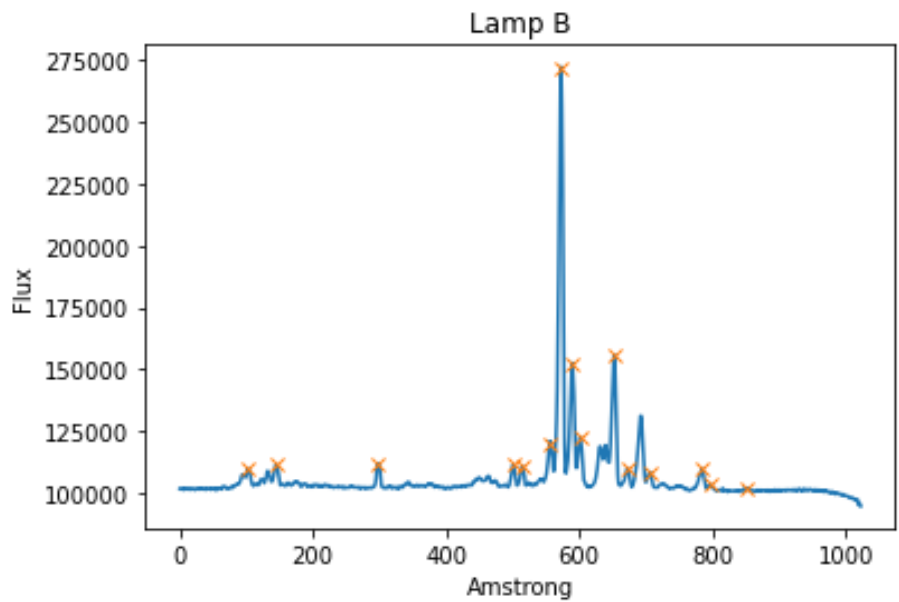

**Figura 3.4:** Espectro de la lámpara de calibración HgAr.

## 3.2 Cámaras CCD

Básicamente un CCD consiste en una matriz de píxeles que recogen los fotones transformando la cantidad de luz en una corriente proporcional de electrones. Esta corriente se amplifica y se convierte a tensión para finalmente, visualizar una imagen digital bidimensional en la pantalla del ordenador.

Al tratarse de instrumentos electrónicos están sujetas al ruido electrónico, pequeñas diferencias entre píxeles y efectos ópticos del instrumento como pueden ser manchas, viñeteado, etc.. [5]

### 3.2.1 Calibración de imágenes digitales

Las cámaras CCD como sistema electrónico que son, tienen un ruido de fondo a causa de la agitación térmica producida por el movimiento de electrones. Esto se traduce en errores de medida en cuanto a la luz que el equipo capta de los cuerpos celestes, se trata por tanto de una información extra que deteriora las imágenes.

El dark, denominado así por ser el promedio de las imágenes que se toman tapando el objetivo del telescopio de manera que no llegue ninguna luz exterior al chip de la CCD, sirve para corregir el ruido anteriormente citado. La temperatura del equipo es muy importante durante este proceso, por tanto es esencial que el tiempo de exposición de los darks sea el mismo que el de la imagen además de tomarlos uno detrás de otro para asegurarnos que la temperatura no cambia.

Para corregir defectos en el proceso de fabricación se toman los flats.[2]. Al flat promedio que resulta hay que restarle su dark promedio llamando a la imagen resultante *flat master*, la cual nos muestra la variación de la sensibilidad de la CCD en toda su superficie.

Una vez se obtiene el *flat master* del instrumento, el proceso básico de calibración de una imagen digital consiste en restar a dicha imagen el promedio de los darks tomados y posteriormente normalizarla con el *flat master*. La (Figura 3.5), muestra esquemáticamente este proceso de una manera resumida [6].

---

[2]Tomas planas que se obtienen exponiendo el equipo a una iluminación uniforme.

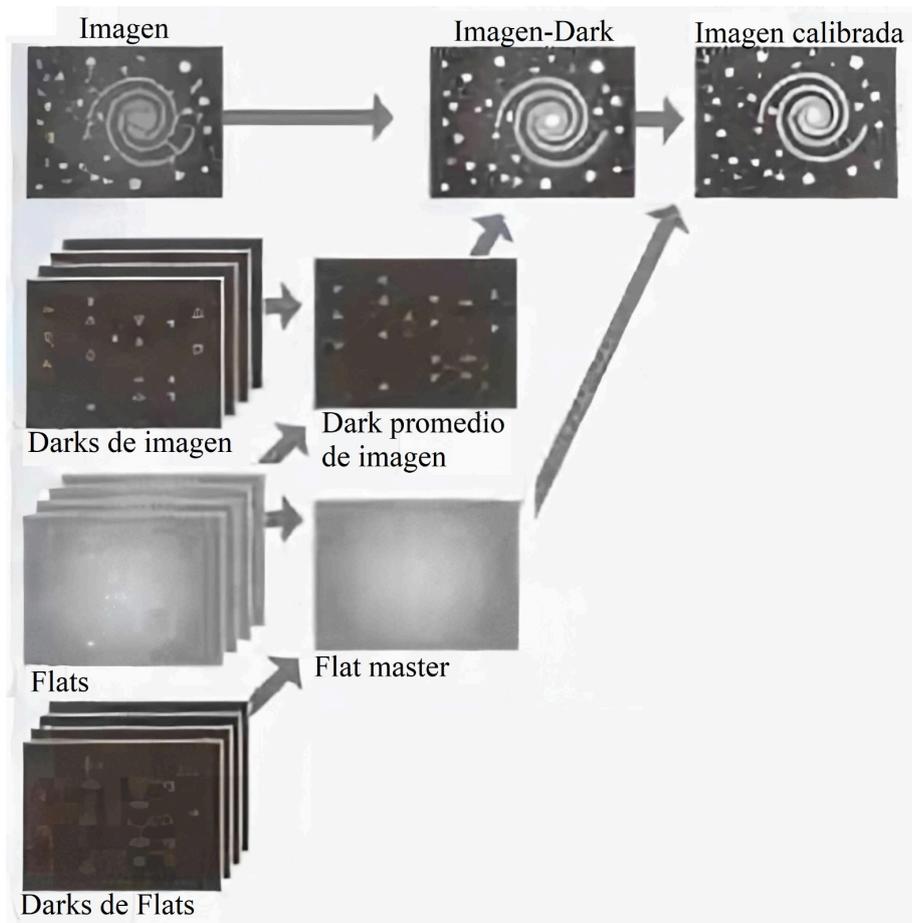

**Figura 3.5:** Esquema de calibración de una imagen digital [7]

# 4 Adquisición de imágenes con BOOTES 2

Este capítulo hace hincapié en el proceso paso a paso desde la localización de un objeto astronómico hasta la obtención de una imagen para procesar mediante el software desarrollado.

## 4.1 Stellarium

Una herramienta muy importante a la hora de localizar estrellas o cuerpos celestes para su estudio es Stellarium. Se trata de un software libre que permite simular la bóveda celeste tal y como la vemos nosotros[8].

Entre sus ventajas destacan:

1. Un catálogo que contiene 600.000 estrellas ampliable hasta 177 millones, además de planetas, satélites en movimiento y objetos del espacio profundo.

2. Dispone de un control para la localización (Figura 4.1) y el tiempo (Figura 4.2). De esta forma, se puede comprobar donde se encontrará el objeto que se desea observar desde diferentes ubicaciones y horas.

3. Versátil herramienta de búsqueda para encontrar los objetos mediante su nombre, su posición o mirando en listas (Figura 4.3).

4. Datos útiles de cada cuerpo como sus coordenadas Ascensión Recta (RA) y Declinación (DEC), su magnitud (medida del brillo), altitud, periodo orbital, etc... (Figura 4.4).

5. También simula fenómenos astronómicos, tales como lluvias de meteoros, eclipses lunares y solares.

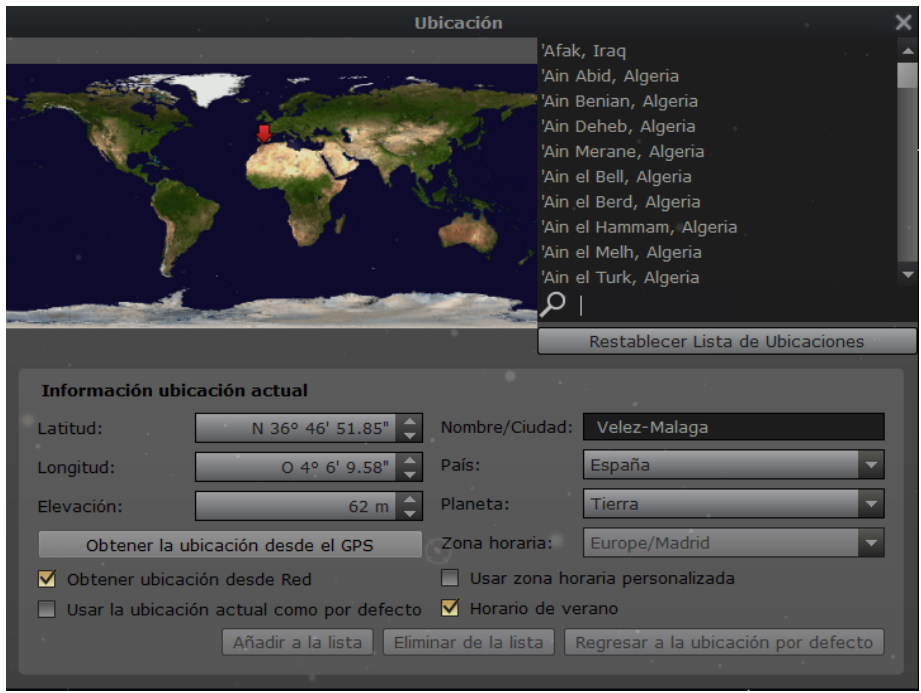

**Figura 4.1:** Ventana de ubicación de Stellarium.

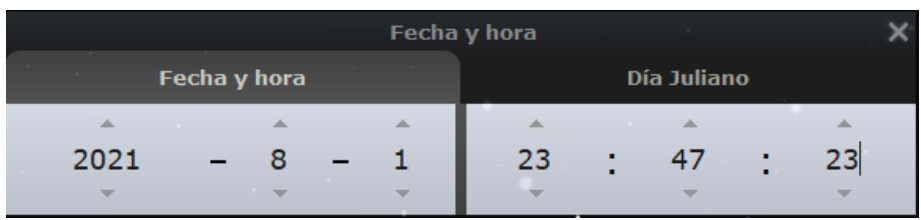

**Figura 4.2:** Ventana de Fecha/Hora de Stellarium.

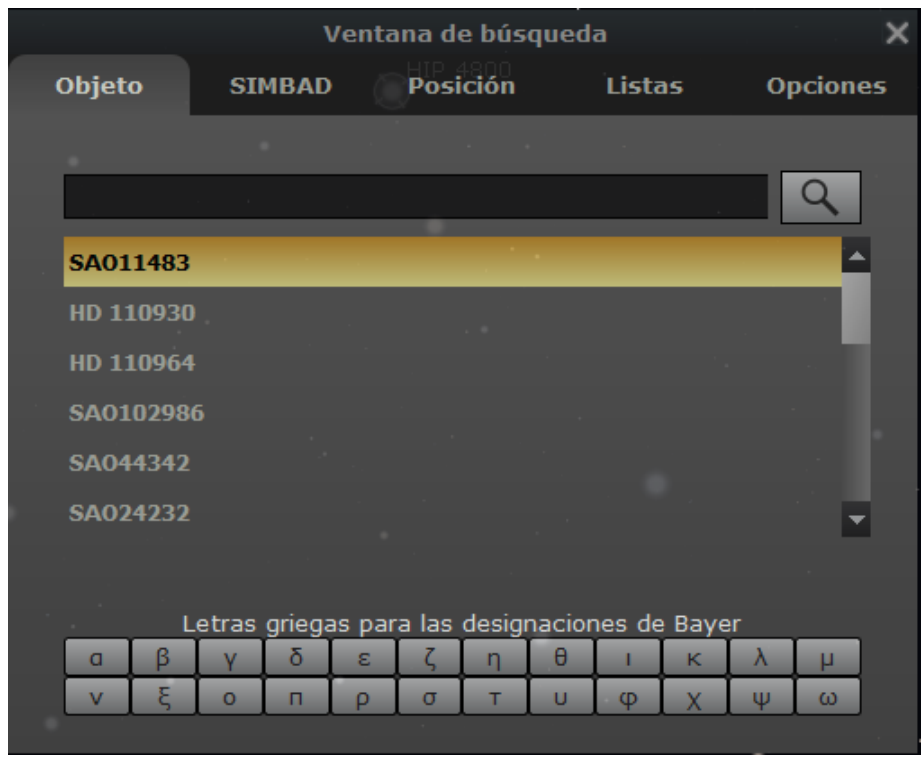

**Figura 4.3:** Ventana de búsqueda de Stellarium.

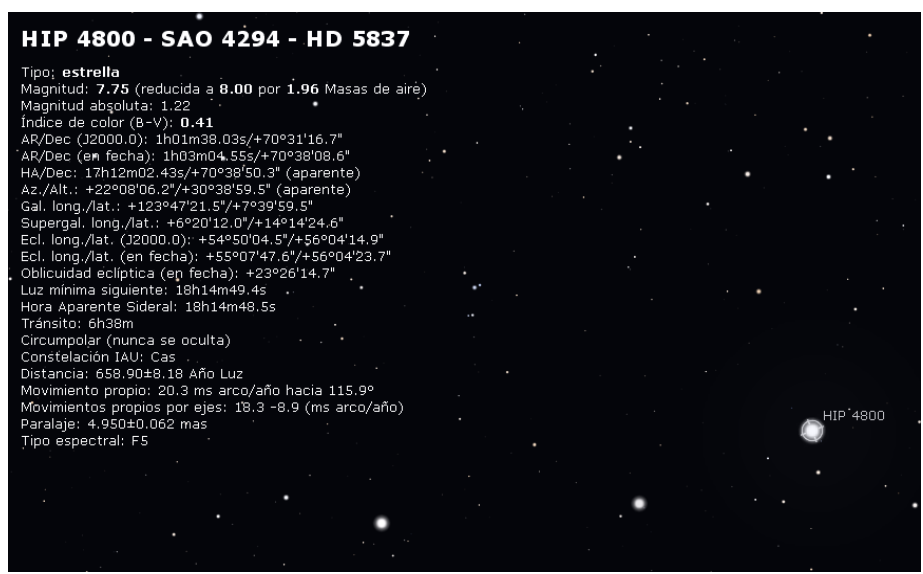

**Figura 4.4:** Datos de HIP 4800 aportados por el software.

Una vez localizada la estrella se busca con el telescopio, pero antes hay que comprobar una serie de requisitos para que sea observable por el telescopio tales como:

1. Las condiciones meteorológicas para la observación deben ser las adecuadas: Nivel apropiado de partículas en suspensión, densidad baja de nubes, viento reducido (aunque en el caso de BOOTES 2 al disponer de cúpula no es un gran inconveniente).

2. El objeto se tiene que encontrar un mínimo de 20 grados sobre el horizonte.

3. La magnitud no puede ser muy baja puesto que el espectrógrafo tendría problemas para captarla.

## 4.2 BOOTES 2 Executor

En este apartado se explicará como operar con el telescopio para introducir los datos conseguidos trabajando con Stellarium y a partir de ellos obtener una imagen.

Primeramente se ingresa a la web del telescopio. Después de identificarse (Figura 4.5), se accede a la pestaña *Executor*. En esta ventana se encuentra la interfaz de usuario con la que se dirigirá la observación (Figura 4.6).

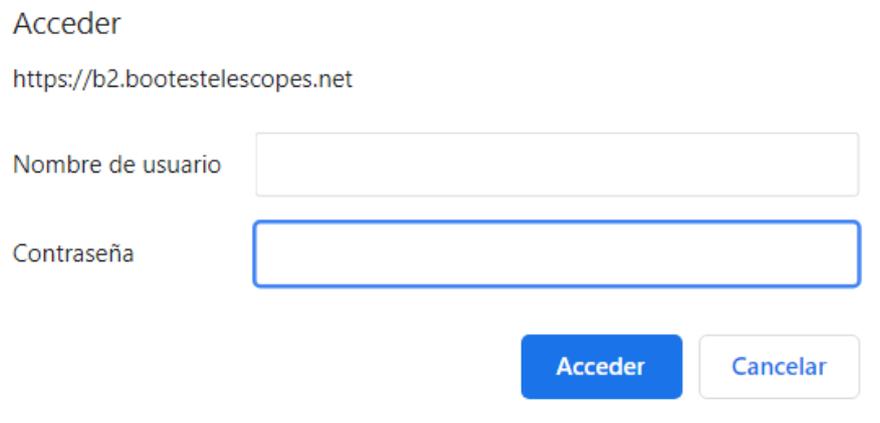

**Figura 4.5:** Ventana de identificación de usuario.

**Figura 4.6:** Interfaz de usuario del Executor.

A continuación se describirán brevemente los campos a rellenar para solicitar una observación:

1. **Target Name:** Nombre que se le quiere dar a la observación. Se incluirá en el fichero generado.

2. **Priority:** Como su propio nombre indica marca la prioridad de la observación, por defecto se encuentra en la más baja (100), pero puede aumentar hasta 20.

3. **RA HH:MM:SS.SS (J2000):** Coordenadas de RA.

4. **DEC +DD:MM:SS (J2000):** Coordenadas de DEC.

5. **CCD:** En el caso de BOOTES 2 al solo disponer de una CCD para elegir se deja a 0.

6. **Exp. Time:** Tiempo de exposición de las imágenes en segundos. Si el tiempo es menor de 7 segundos la imagen no incluirá astrometría.

7. **Filter Wheel/Autofocus:** Da a elegir entre los filtros disponibles en el telescopio. Para este proyecto se utilizará el filtro *Slit 4.3* a la hora de tomar espectros.

8. **Starting UT Date y Time:** Selecciona la fecha y hora de la observación en formato (HH:MM:SS) y en Tiempo Universal Coordinado (UTC).

9. **Not After UT Date y Time:** Selecciona la fecha y hora límite para tomar la observación antes de descartarla en formato (HH:MM:SS) y en UTC.

10. **Schedule:** Dispone de tres opciones: tan pronto como sea posible (ASAP), tan bien como sea posible (AWAP) y la opción por defecto (FIXED) que planificará la observación con la fecha y hora propuestas en los campos anteriores.

Una vez se sabe como rellenar los campos se manda una primera observación con un tiempo de exposición entre 20 y 30 segundos y el filtro *Open* para observar la estrella.

Las coordenadas que se cogen de Stellarium son aproximadas, pero no exactas. Por tanto, se utiliza la herramienta de la web del telescopio *Transient detection* para visualizar las coordenadas exactas de la estrella que queremos observar comparando su magnitud con la que ofrece Stellarium (Figuras 4.7 y 4.8).

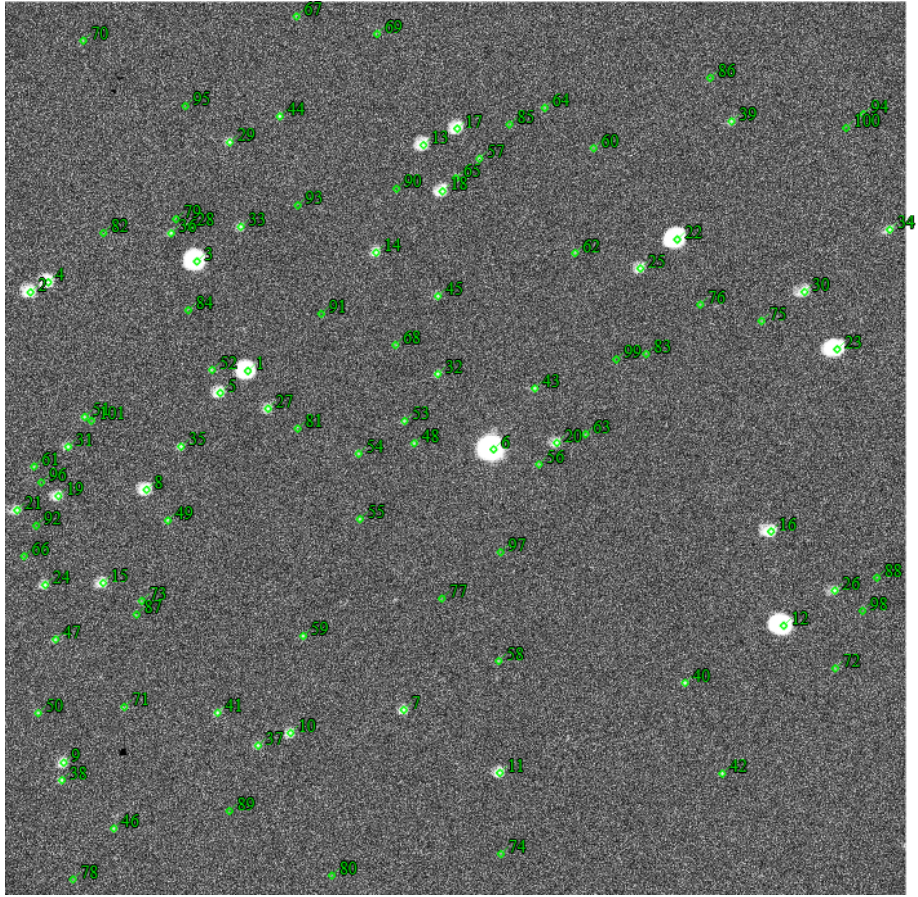

**Figura 4.7:** FITS de Transient Detection de SAO 22966.

| Log | Total: | 101 | MaxMag: | 16.673 | Filter: | open | FIT URL | | |
|---|---|---|---|---|---|---|---|---|---|
| N# | XPixel | YPixel | RA | DEC | Measured Mag | Catalog Mag | Mag Error | Detection | NEO & Comet Checker |
| 1 | 277 | 423 | 02:10:29.37 (32.6224) | +57:54:39.4 (57.9109) | 10.451 [0.013] | 10.86 | -0.409 | | |
| 2 | 30 | 333 | 02:10:22.38 (32.5932) | +57:52:12.49 (57.8701) | 12.383 [0.014] | 12.603 | -0.22 | | |
| 3 | 219 | 298 | 02:10:19.9 (32.5829) | +57:54:05.81 (57.9016) | 9.934 [0.014] | 10.151 | -0.217 | | |
| 4 | 50 | 322 | 02:10:21.53 (32.5897) | +57:52:24.62 (57.8735) | 12.069 [0.014] | 12.385 | -0.316 | | |
| 5 | 246 | 448 | 02:10:31.23 (32.6301) | +57:54:20.43 (57.9057) | 12.703 [0.013] | 13.021 | -0.318 | | |
| 6 | 556 | 513 | 02:10:36.39 (32.6516) | +57:57:25.51 (57.9571) | 8.701 [0.013] | 9.037 | -0.336 | | |
| 7 | 454 | 811 | 02:10:58.71 (32.7446) | +57:56:21.97 (57.9394) | 13.272 [0.014] | 13.509 | -0.238 | | |
| 8 | 162 | 559 | 02:10:39.46 (32.6644) | +57:53:29.46 (57.8915) | 12.554 [0.014] | 12.696 | -0.142 | | |
| 9 | 68 | 871 | 02:11:02.82 (32.7617) | +57:52:30.42 (57.8751) | 13.227 [0.013] | 13.444 | -0.217 | | |
| 10 | 326 | 837 | 02:11:00.55 (32.7523) | +57:55:04.84 (57.918) | 13.42 [0.015] | 13.603 | -0.183 | | |
| 11 | 563 | 883 | 02:11:04.26 (32.7677) | +57:57:26.54 (57.9574) | 13.025 [0.014] | 13.193 | -0.168 | | |
| 12 | 886 | 714 | 02:10:51.92 (32.7163) | +58:00:40.78 (58.0113) | 9.809 [0.013] | 9.981 | -0.172 | | |
| 13 | 477 | 165 | 02:10:10.13 (32.5422) | +57:56:40.62 (57.9446) | 12.622 [0.015] | 12.685 | -0.063 | | |
| 14 | 423 | 288 | 02:10:19.33 (32.5805) | +57:56:07.57 (57.9354) | 13.443 [0.015] | 13.575 | -0.133 | | |
| 15 | 113 | 665 | 02:10:47.42 (32.6976) | +57:52:59.04 (57.8831) | 13.091 [0.014] | 13.204 | -0.113 | | |
| 16 | 871 | 607 | 02:10:43.79 (32.6824) | +58:00:33.2 (58.0092) | 12.404 [0.014] | 12.561 | -0.157 | | |
| 17 | 515 | 146 | 02:10:08.75 (32.5364) | +57:57:03.71 (57.951) | 12.527 [0.014] | 12.64 | -0.113 | | |
| 18 | 498 | 218 | 02:10:14.13 (32.5589) | +57:56:53.14 (57.9481) | 12.722 [0.014] | 12.816 | -0.095 | | |

**Figura 4.8:** Tabla de Transient Detection de SAO 22966.

Este paso es importante puesto que se necesita que la estrella esté centrada en la imagen para que cuando se ponga la rendija se pueda tomar el espectro. Por este motivo, se vuelve a enviar una nueva observación con las coordenadas que nos devuelve el *Transient detection* para ver si se ha conseguido el objetivo (Figura 4.9).

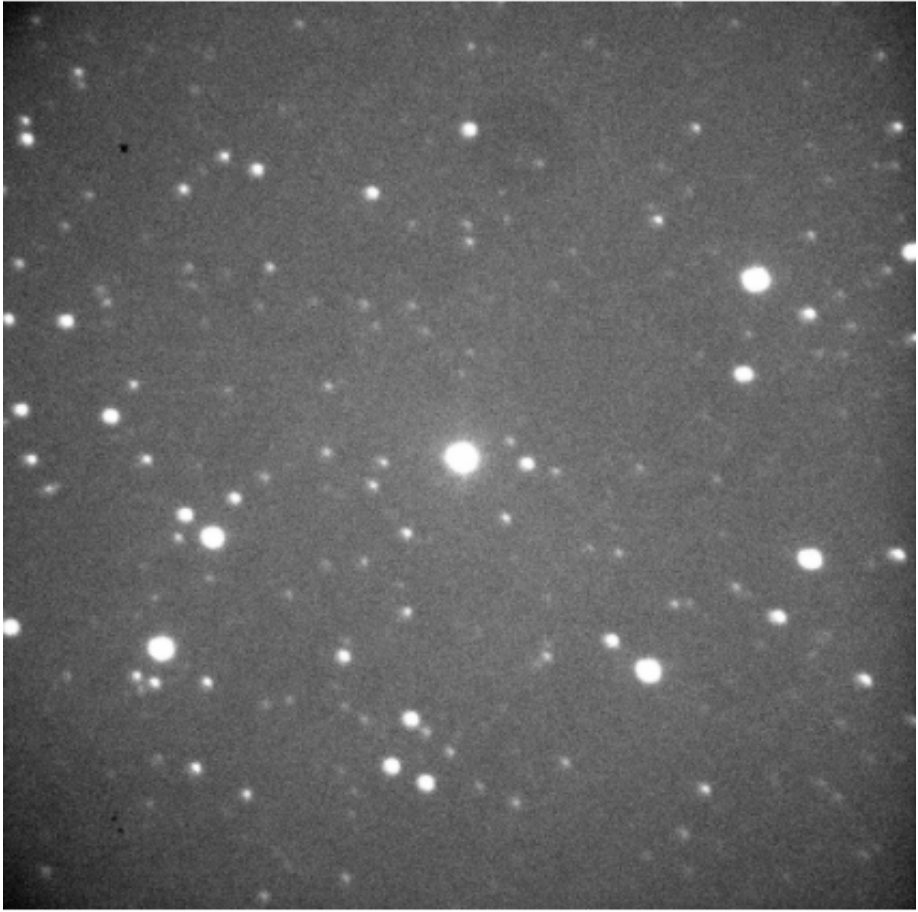

**Figura 4.9:** Estrella SAO 22966 centrada en la imagen.

Tras comprobar que la estrella está centrada en la imagen se procede al último paso. Se vuelven a introducir las coordenadas, pero en este caso se cambian algunos parámetros. Se elige un tiempo de exposición de 300s (tiempo máximo permitido) para observar el espectro y además se elige como filtro la rendija *Slit 4.3*, puesto que es la que más luz deja pasar y dará un espectro más sencillo de analizar.

Finalmente, se pulsa el botón *Colores Spectrum* en el panel y se lanzan automáticamente tres observaciones (Figura 4.10). Dos con tiempo de exposición de 15s y filtro *Open* para comprobar que el telescopio se encuentra enfocando al objetivo y una última con la rendija ya puesta y tiempo de exposición de 300s para tomar el espectro (Figura 4.11).

| 562578 | SLITSAO22966 | 21 | 32,6517 | 57,95 | 300 | Slit 4.3 | 28/08/2021 | 22:09:00 | 31/08/2021 | 17:00:00 | OK | Download | Download | R/E Copy Simbad |
| 562577 | POINTSAO22966 | 21 | 32,6517 | 57,95 | 15 | Open | 28/08/2021 | 22:09:00 | 31/08/2021 | 17:00:00 | OK | Download | Download | R/E Copy Simbad |
| 562576 | POINTSAO22966 | 21 | 32,6517 | 57,95 | 15 | Open | 28/08/2021 | 22:09:00 | 31/08/2021 | 17:00:00 | OK | Download | Download | R/E Copy Simbad |

**Figura 4.10:** Secuencia de envío de observaciones para hallar el espectro.

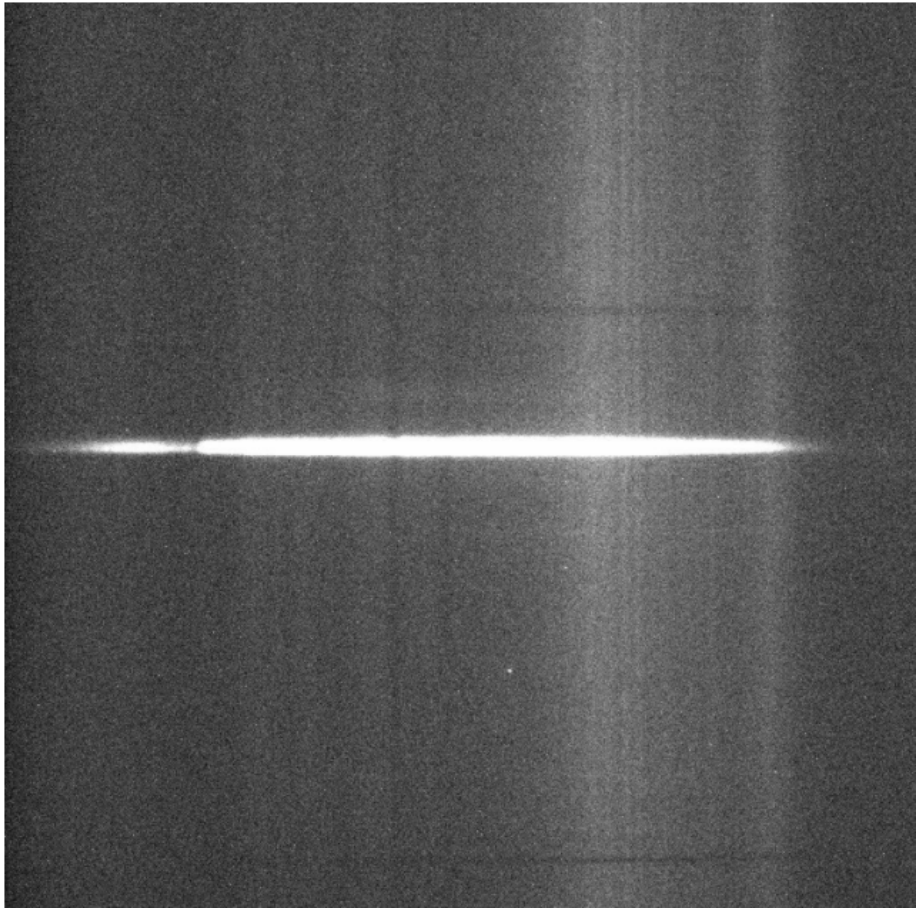

**Figura 4.11:** Espectro en 2 dimensiones de la estrella SAO 22966.

# 5 Programa desarrollado

Tras describir como operar el telescopio para conseguir las imágenes con las que se va a trabajar, en este capítulo se explicará tanto el software desarrollado para el procesamiento de los archivos Flexible Image Transport System (FITS) y la obtención de su espectro, como las librerías y las herramientas matemáticas utilizadas.

Se trabajará en Spyder que es un entorno científico gratuito y de código abierto escrito en Python y diseñado para científicos e ingenieros. Cabe destacar que los comentarios realizados en el código se encuentran en Inglés para que cualquier persona que acceda a él para futuras implementaciones pueda entenderlo.

Añadir además, que el código desarrollado se encuentra ubicado en el siguiente repositorio: https://github.com/spaceuma/BOOTES-COLORES

Este repositorio se encuentra cerrado, y para su acceso se requiere solicitar permisos a Carlos Jesús Pérez del Pulgar Mancebo en carlosperez@uma.es

## 5.1 Paquete Astropy

El proyecto Astropy comienza como un esfuerzo colectivo de la comunidad científica para desarrollar un paquete básico estandarizado para la Astronomía en Python. [9].

Sus librerías dotan al usuario de las herramientas necesarias para operar con imágenes astronómicas. Para el desarrollo de este TFG, se han utilizado algunas de ellas.

### 5.1.1 FITS

El formato FITS es el utilizado más comúnmente en el mundo astronómico y nace de la necesidad de almacenar toda la información contenida en una misma observación. Con el avance tecnológico, los telescopios son capaces de recoger una gran cantidad de datos, no solo imágenes sino también espectros electromagnéticos, listas de fotones, coordenadas, etc... Además, hacía falta un formato estandarizado para que las distintas agencias e instituciones científicas a lo largo de todo el planeta presentaran esta información [10].

Como ventaja principal del formato FITS se destaca la cabecera que además es legible en American Standard Code for Information Interchange (ASCII). Cada archivo FITS consiste en una o más cabeceras que contienen secuencias de 80 cadenas de caracteres fijos. Estas contienen información muy útil a la hora de procesar o entender la procedencia de la imagen puesto que contienen fecha y hora de la captura, temperatura, coordenadas y además la astrometría de la imagen. (Figura 5.1).

```
SIMPLE  =                    T
BITPIX  =                   16 /8 unsigned int, 16 & 32 int, -32 & -64 real
NAXIS   =                    2 /number of axes
NAXIS1  =                 1024 /fastest changing axis
NAXIS2  =                 1024 /next to fastest changing axis
BSCALE  =    1.0000000000000000 /physical = BZERO + BSCALE*array_value
BZERO   =   32768.000000000000 /physical = BZERO + BSCALE*array_value
DATE-OBS= '2021-05-05T23:48:31.488'
EXPTIME =   300.00000000000000 /Exposure time in seconds
EXPOSURE=   300.00000000000000 /Exposure time in seconds
SET-TEMP=  -50.000000000000000 /CCD temperature setpoint in C
CCD-TEMP=  -50.041999816894531 /CCD temperature at start of exposure in C
XPIXSZ  =    13.000000000000000 /Pixel Width in microns (after binning)
YPIXSZ  =    13.000000000000000 /Pixel Height in microns (after binning)
XBINNING=                    1 /Binning factor in width
YBINNING=                    1 /Binning factor in height
XORGSUBF=                    0 /Subframe X position in binned pixels
YORGSUBF=                    0 /Subframe Y position in binned pixels
FILTER  = 'Slit 4.3' /          Filter used when taking image
IMAGETYP= 'Light Frame' /       Type of image
FOCALLEN=    10000.000000000000 /Focal length of telescope in mm
APTDIA  =    600.00000000000000 /Aperture diameter of telescope in mm
APTAREA =    212057.51001834869 /Aperture area of telescope in mm^2
SBSTDVER= 'SBFITSEXT Version 1.0' /Version of SBFITSEXT standard in effect
SWCREATE= 'MaxIm DL Version 6.16 181226 0V2FC' /Name of software
SWSERIAL= '0V2FC-9AT7K-7RQQA-VM6Y5-N6V9X-7J' /Software serial number
SITELAT = '36 45 33' /          Latitude of the imaging location
SITELONG= '-04 02 28' /         Longitude of the imaging location
JD      =    2459340.4920262732 /Julian Date at start of exposure
JD-HELIO=    2459340.4896715917 /Heliocentric Julian Date at exposure midpoint
OBJECT  = 'SLITF   '
TELESCOP= '        ' /          telescope used to acquire this image
INSTRUME= 'Andor Tech'
OBSERVER= '       '
NOTES   = '       '
FLIPSTAT= '       '
MAXIMDT = '2021-05-05T23:48:31.07'
GPSTIME = '23:48:31.488'
OBJERA  =    112.22166666666701
OBJEDEC =    87.381977777777806
OBJEALT =    35.917519216796684
OBJEAZ  =    356.76108680682557
OBSLAT  = '36,7591666666667'
OBSLONG = '-4,04111111111111'
OBJHEIG = '50      '
SWOWNER = 'IAA-CSIC-1 (Alberto J Castro-Tirado)' /Licensed owner of software
END
```

**Figura 5.1:** Cabecera de un archivo FITS.

El uso de esta librería ha permitido trabajar con las imágenes descargadas del servi-dor del telescopio y operar con ellas.

### 5.1.2 NumPy

La librería NumPy está especializada en el cálculo numérico y el análisis de datos, especialmente de gran volumen. Además, la característica más importante de la que dispone para este proyecto en cuestión es su estructura de datos *ndarray* que permite operar con matrices de $n$ dimensiones[11].

Todos los elementos dentro de una matriz deben ser del mismo tipo, lo cual encaja perfectamente con como se extraen los datos de una imagen. Se trabajará con arrays de 1024x1024 en los que se guardan la información de cada píxel de la imagen (Figura 5.2).

**Figura 5.2:** NumPy array de datos de una imagen.

### 5.1.3 Matplotlib

La biblioteca Matplotlib permite al usuario la generación de gráficos a partir de datos guardados en listas o arrays, es por esto que es un complemento perfecto a la librería NumPy anteriormente citada y permite ver de una forma más descriptiva los distintos pasos que realiza el software [12].

De esta forma, a partir de la imagen obtenida del telescopio y de su procesamiento se podrán extraer gráficas que muestren distintos resultados. Un ejemplo podría ser el de la (Figura 5.3), que muestra el FITS con distintas escalas de colores.

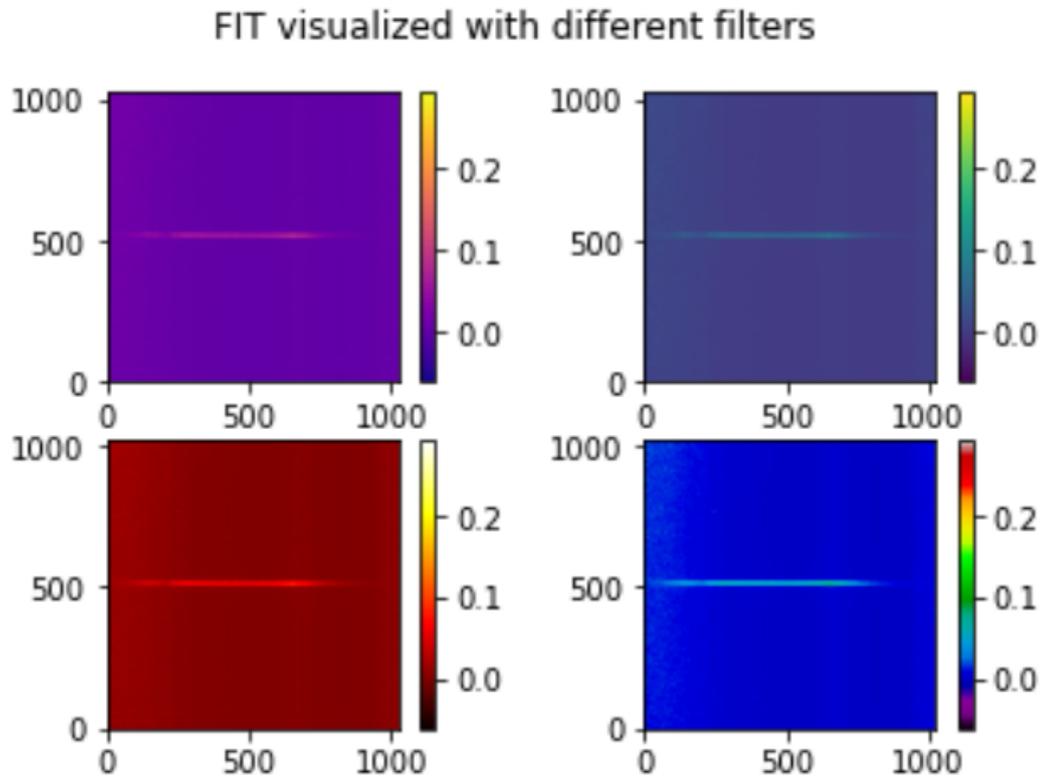

**Figura 5.3:** FIT visualizado con distintas escalas de colores.

## 5.2 Otras funcionalidades

Durante la realización de este TFG se han desarrollado diversas soluciones que no se han implementado finalmente en el código, pero que merece la pena comentar.

### 5.2.1 Mostrar la cabecera y otros datos relevantes

Es posible una vez abierto el archivo FITS dentro de Spyder, mostrar en la consola tanto su cabecera (Figura 5.4) como algunos datos que pueden ser de interés a la hora de analiza la imagen (Figura 5.5) como el tipo de dato, la desviación estándar o el tamaño.

```
---------------------------- Header ----------------------------
SIMPLE  =                    T
BITPIX  =                   16 /8 unsigned int, 16 & 32 int, -32 & -64 real
NAXIS   =                    2 /number of axes
NAXIS1  =                 1024 /fastest changing axis
NAXIS2  =                 1024 /next to fastest changing axis
BSCALE  =    1.0000000000000000 /physical = BZERO + BSCALE*array_value
BZERO   =    32768.000000000000 /physical = BZERO + BSCALE*array_value
DATE-OBS= '2021-08-28T22:13:41.825'
EXPTIME =     300.00000000000000 /Exposure time in seconds
EXPOSURE=     300.00000000000000 /Exposure time in seconds
SET-TEMP=    -50.000000000000000 /CCD temperature setpoint in C
CCD-TEMP=    -50.041999816894531 /CCD temperature at start of exposure in C
XPIXSZ  =     13.000000000000000 /Pixel Width in microns (after binning)
YPIXSZ  =     13.000000000000000 /Pixel Height in microns (after binning)
XBINNING=                    1 /Binning factor in width
YBINNING=                    1 /Binning factor in height
XORGSUBF=                    0 /Subframe X position in binned pixels
YORGSUBF=                    0 /Subframe Y position in binned pixels
FILTER  = 'Slit 4.3' /         Filter used when taking image
IMAGETYP= 'Light Frame' /       Type of image
FOCALLEN=    10000.000000000000 /Focal length of telescope in mm
APTDIA  =     600.00000000000000 /Aperture diameter of telescope in mm
APTAREA =    212057.51001834869 /Aperture area of telescope in mm^2
SBSTDVER= 'SBFITSEXT Version 1.0' /Version of SBFITSEXT standard in effect
SWCREATE= 'MaxIm DL Version 6.16 181226 0V2FC' /Name of software
SWSERIAL= '0V2FC-9AT7K-7RQQA-VM6Y5-N6V9X-7J' /Software serial number
SITELAT = '36 45 33' /          Latitude of the imaging location
SITELONG= '-04 02 28' /         Longitude of the imaging location
JD      =    2459455.4261755785 /Julian Date at start of exposure
JD-HELIO=    2459455.4332206077 /Heliocentric Julian Date at exposure midpoint
OBJECT  = 'SLITSA034810'
```

**Figura 5.4:** Cabecera representada en la consola.

```
----- Statistics values -----
Min : -14.0
Max : 600.0
Mean : 9.3372163772583
Stdev : 24.596034801570468
Data Type : float64
Image length : 1024
Shape : (1024, 1024)
```

**Figura 5.5:** Algunos datos del FITS

Código 5.1: Mostrar cabecera y datos del FITS

```python
#I show the header
print(f'Image Data Type : {type(image_data)} - Shape : {image_data.shape}\n')
print('----------------------------- Header -----------------------------')
print(repr(header))
print('--------------------------- End Header ---------------------------')

#Some values of the image
print('----- Statistics values -----')
print('Min :', np.min(image_data))
print('Max :', np.max(image_data))
print('Mean :', np.mean(image_data))
print('Stdev :', np.std(image_data)) #Standard deviation
print('Data Type :', image_data.dtype)
print('Image length : ', len(image_data)) #Size list
print('Shape :', image_data.shape) #Dimensions of the array
```

### 5.2.2 Diferentes gráficas del espectro

Otra de las funcionalidades que merece la pena destacar es la de representar gráficamente diferentes datos de la imagen espectral. En este caso, se muestran su magnitud, fase y ángulo (Figura 5.6).

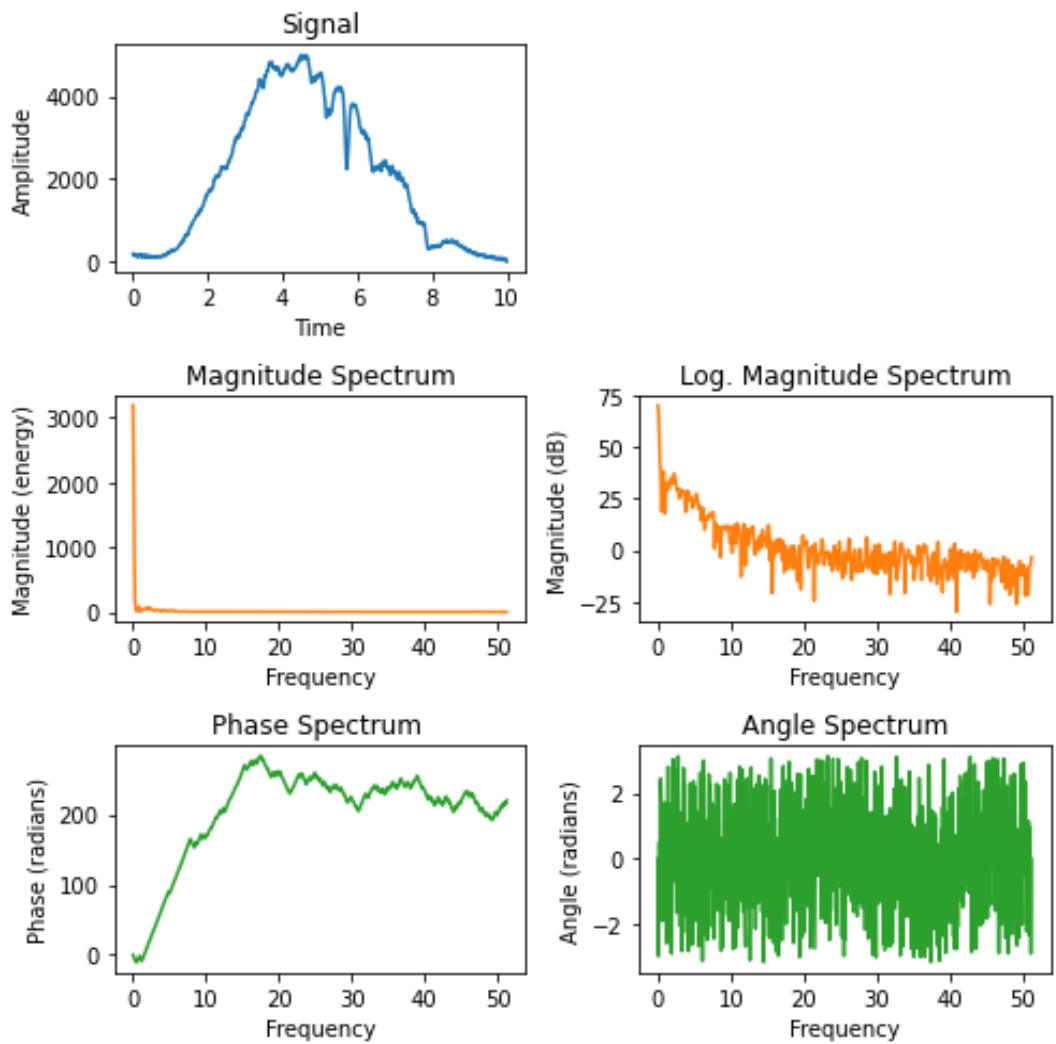

**Figura 5.6:** Gráficas con datos del espectro.

Código 5.2: Representación de datos del espectro

```
#Display different spectrum data
dt = 10/len(spectra) #sampling interval (1024)
Fs = 1 / dt #sampling frequency
t = np.arange(0, 10, dt)

fig, axes = plt.subplots(nrows=3, ncols=2, figsize=(7, 7))

#Display time signal:
axes[0, 0].set_title("Signal")
axes[0, 0].plot(t, spectra, color='C0')
axes[0, 0].set_xlabel("Time")
axes[0, 0].set_ylabel("Amplitude")

#Display different type of spectrum:
axes[1, 0].set_title("Magnitude Spectrum")
```

```
16 axes[1, 0].magnitude_spectrum(spectra, Fs, color='C1')
17
18 axes[1, 1].set_title("Log. Magnitude Spectrum")
19 axes[1, 1].magnitude_spectrum(spectra, Fs, scale='dB', color='C1')
20
21 axes[2, 0].set_title("Phase Spectrum ")
22 axes[2, 0].phase_spectrum(spectra, Fs, color='C2')
23
24 axes[2, 1].set_title("Angle Spectrum")
25 axes[2, 1].angle_spectrum(spectra, Fs, color='C2')
26
27 axes[0, 1].remove() #Don't display empty ax
28
29 fig.tight_layout()
30 plt.show()
```

## 5.3 Representación del espectro

Se empieza este apartado explicando de forma generalizada la secuencia de código que se desarrolla.

La programación se desarrolla en 2 scripts diferenciados. El primero de ellos llamado *spectrum.py* mostrará la gráfica del espectro mientras que el script *calibration.py* se encargará como su propio nombre indica únicamente de la parte de la calibración mediante el uso de la lampara de HgAr.

Cabe destacar que estos dos scripts se han simplificado al máximo para que simplemente hagan su función y no introducir un código muy pesado en el servidor. Sin embargo, a lo largo de la realización de este TFG se ha utilizado un código más minucioso que muestra los resultados de cada paso y que es el que se explicará en esta sección.

### 5.3.1 spectrum.py

**Definición de la función:** La primera línea de código trata de definir la función, para que una vez implementado el programa en el servidor del telescopio genere la gráfica con el espectro.

La función cuenta con tres parámetros:

1. *fits path:* Indica la ruta donde se encuentra ubicada dentro del equipo la imagen FITS.

2. *config path:* Se encarga de introducir el fichero *.txt* creado por textitcalibration.py y que contiene la información de la lampara de calibración.

3. *graph path:* Parámetro de salida que guardará un archivo con la gráfica del espectro.

Código 5.3: Definición de la función spectrum

```
1    def spectrum(fits_path,config_path,graph_path):
```

**Librerías:** Se definen las librerías que se van a utilizar a lo largo del programa.

Código 5.4: Librerías utilizadas

```
1    #Libraries I need to use
2    from astropy.io import fits
3    from scipy import ndimage
4    import numpy as np
5    import matplotlib.pyplot as plt
6    from scipy.signal import find_peaks
7    from scipy.optimize import curve_fit
```

**Apertura y conversión del FITS:** El siguiente paso es abrir la imagen con la que se va a trabajar desde el directorio donde se encuentra. Una vez hecho esto se trabaja con el FITS como una lista y por último, se guardan las dimensiones de esa lista. En el caso de BOOTES 2 las imágenes son de 1024x1024 píxeles como se puede ver en la pestaña *Variable explorer* de Spyder (Figura 5.7).

| Name | Type | Size | Value |
|---|---|---|---|
| fit_open | io.fits.hdu.hdulist.HDUList | 1 | HDUList object of astropy.io.fits.hdu.hdulist module |
| hdu_list | io.fits.hdu.hdulist.HDUList | 1 | HDUList object of astropy.io.fits.hdu.hdulist module |
| header | io.fits.header.Header | 1 | Header object of astropy.io.fits.header module |
| image_size | tuple | 2 | (1024, 1024) |

**Figura 5.7:** Ventana Variable explorer de Spyder.

Código 5.5: Apertura y uso del FITS como lista

```
1    #To working with the FIT, first of all open the directory
2    fit_open = fits.open(fits_path)
3    #I work with the .fit as a list
4    hdu_list = fit_open
5    hdu_list.info()
6    #I save dimensions of the array
7    image_size = hdu_list[0].shape
```

**Calibración de la imagen:** En esta parte del código se procede a conseguir que el espectro se vea de la forma más *"limpia"* posible mediante el uso de distintos métodos de calibración. En primer lugar, se convierten los datos a tipo float para evitar problemas. A continuación se le realiza a la imagen un filtro de la mediana para evitar los puntos calientes[1] que puede producir el instrumento.

Seguidamente, se abre el bias que estará ubicado en una carpeta fija dentro del servidor y se le resta a la imagen calibrada, para después darle la vuelta ya que el telescopio toma la imagen girada.

---

[1] Píxeles cuyo brillo crece linealmente con el tiempo de exposición.

A pesar de haber calculado el flatmaster no se utilizará, puesto que genera más distorsión en la imagen en lugar de corregirla. Finalmente, se mostrará la imagen corregida (Figura 5.8).

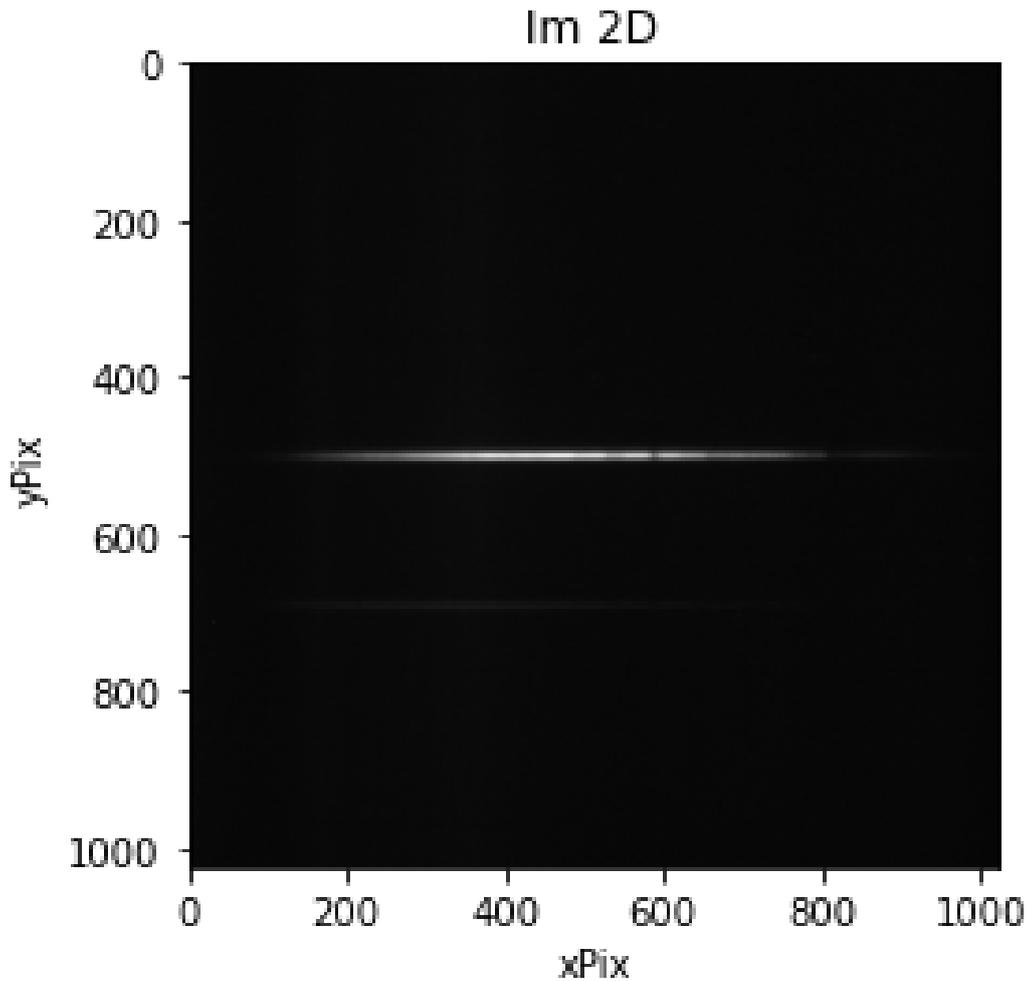

**Figura 5.8:** Imagen del espectro tras el proceso de calibración.

Código 5.6: Calibración del espectro

```
1  #Turn data to float, in order to avoid overflow
2  image_data_ori = hdu_list[0].data*1.0
3  #Median filter with a 2-pixel window, avoids hot spots
4  image_data = ndimage.median_filter(image_data_ori,(2))
5  #Open the bias
6  bias = fits.open(r'D:\Escritorio\GIERM\TFG\COLORES\masterbias.fits')
```

```
7   #Get the image data subtracting the bias and rotating the result
8   image_data = np.flip(image_data−bias[0].data)
9
10  #Plot the image
11  plt.figure(1)
12  plt.imshow(image_data, cmap='gray')
13  plt.xlabel('xPix')
14  plt.ylabel('yPix')
15  plt.title('Im 2D')
```

**Primera aproximación a la gráfica del espectro:** Una vez se consigue una imagen calibrada se procede a extraer la gráfica de su espectro que es la finalidad de este proyecto. Para ello se realiza el sumatorio de todos los píxeles por columnas y se representa su resultado (Figura 5.9).

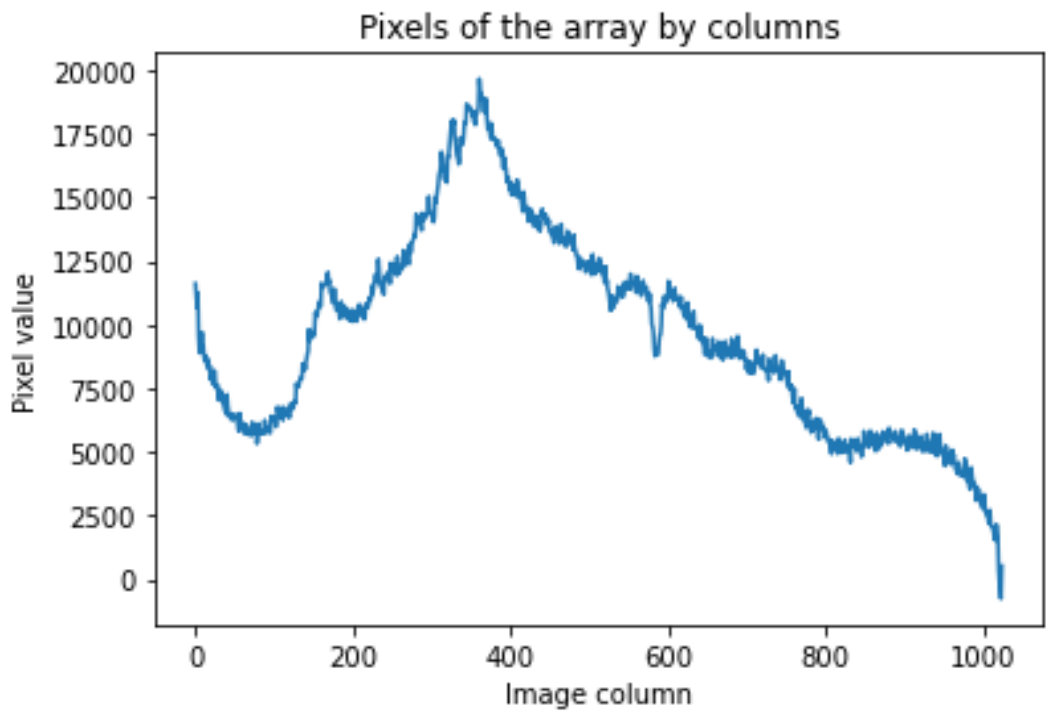

**Figura 5.9:** Sumatorio de todos los píxeles por columnas.

Código 5.7: Primera aproximación
```
1 #This sentence print all the pixels of the array by columns to see the
2 #spectrum without filters
3 sumc=sum(image_data)
4 #Start figure
5 plt.figure(2)
6 plt.xlabel('Image column')
7 plt.ylabel('Pixel value')
8 plt.title('Pixels of the array by columns')
9 plt.plot(sumc)
```

```
10 plt.show()
```

El problema que se aprecia, es que para realizar este sumatorio se han utilizado todos los píxeles de la imagen lo cual desvirtúa el resultado. Lo interesante sería utilizar solo la parte de la imagen donde se encuentra el espectro, donde se coloca la rendija.

**Sumatorio de píxeles por filas:** Para localizar la Region of interest (ROI), se operará ahora por filas. El procedimiento, es similar a hacerlo por columnas con la única diferencia de que hay que transponer la imagen antes de hacer el sumatorio.

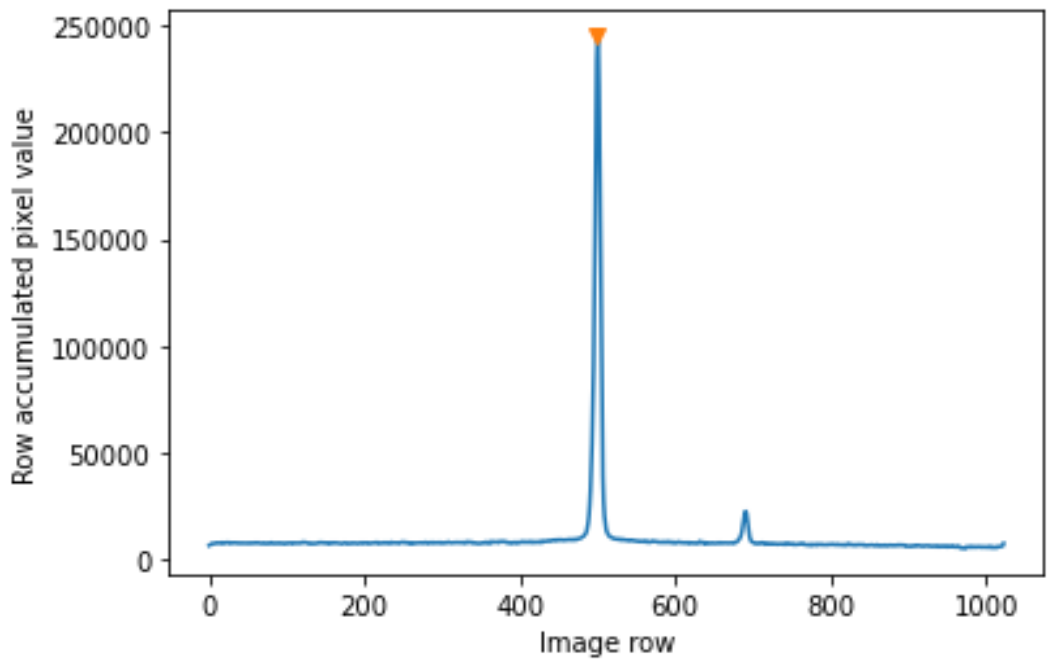

**Figura 5.10:** Imagen de todos los píxeles por filas.

Código 5.8: Sumatorio y representación por filas

```
1 #This sentence print all the pixels of the array by rows
2 #I turn the image around
3 sumf=sum(np.transpose(image_data))
4 #Start figure
5 plt.figure(3)
6 plt.plot(sumf)
7 plt.xlabel('Image row')
8 plt.ylabel('Row accumulated pixel value')
9
10 #I find the the maximum and minimum value for y
11 maxrow=max(sumf)
12 minrow=min(sumf)
13 #I set a threshold of 20% of the value between max and min row value
14 threshold = (maxrow−minrow)*0.2
15
```

```
16 #I put a symbol to mark the maximum
17 localmaxima,_= find_peaks(sumf, distance=threshold)#I don't want 2nd part of the function
18 plt.plot(localmaxima, sumf[localmaxima], "v")
19 plt.show()
```

El resultado muestra (Figura 5.10) una distribución muy similar a una gaussiana. Por tanto, se calcula el máximo valor *(maxrow)* y el mínimo *(minrow)* y se establece un umbral *(threshold)* del 20 por ciento entre el valor máximo y el mínimo.

Por último, se localiza el máximo local en el eje X y se coloca una flecha para marcarlo.

**Definición y representación del ajuste gaussiano:** En esta parte del código se definirá una función que devuelva la expresión de la gaussiana. Se calcula el valor máximo de la función *[b]*, se hace una estimación de la desviación estándar *[c]*, se calcula la amplitud *[a]* y el valor mínimo para alcanzar la función *[bias]*.

Una vez hecho esto, se guarda el nuevo valor ajustado de las desviación estándar *[c]* (Figura 5.11) que se calcula mediante la función *curve fit*. Finalmente, se representa la gráfica con los datos y el ajuste que se realiza (Figura 5.12).

```
----- Standard deviation -----
c: 10
Adjusted c: 3.5629064434988678
```

**Figura 5.11:** Desviación estándar estimada y tras el ajuste.

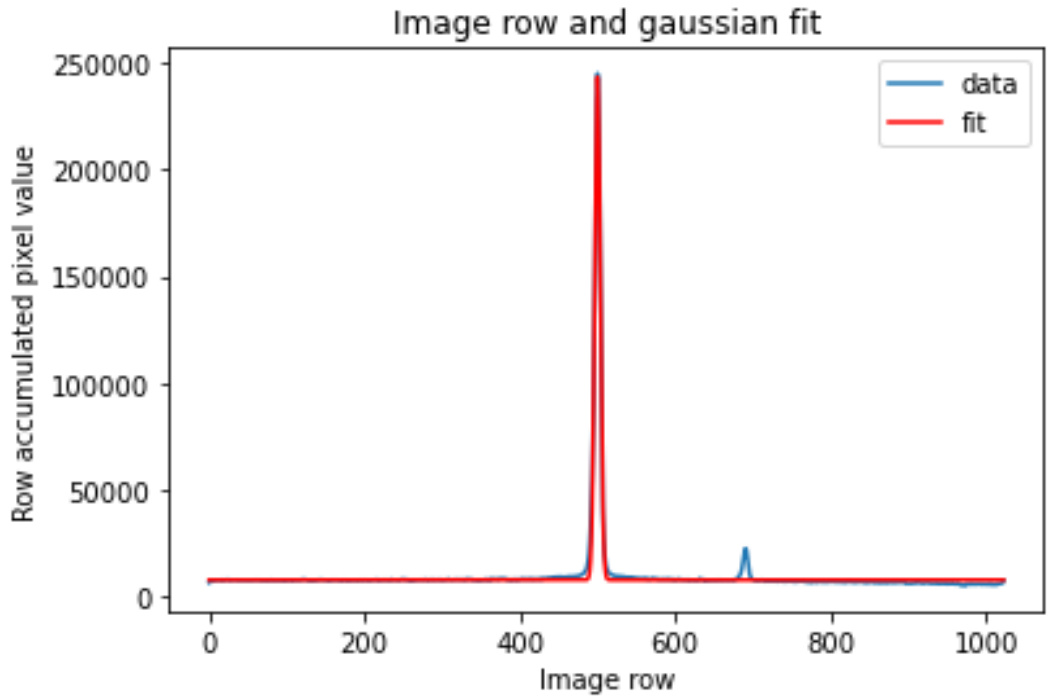

**Figura 5.12:** Ajuste de la gaussiana.

Código 5.9: Ajuste de la gaussiana

```python
def gauss_function(x, a, b, c, bias):
    return bias + a*np.exp(-(x-b)**2/(2*c**2)) #Expression of the gaussian
"""
- I know where has to be the maximum value of the gaussian, in the maximum value [b]
- I make a first estimation of the standard deviation and save its value in [c]
- I also know the amplitude of the Gaussian, max value - min value [a]
- Finally, a Gaussian is centered on the y-axis = 0, so I will have to add
  the minimum value to raise the function to the desired point [bias].
"""
b = int(localmaxima[0])
c = 10 #Started value
a = maxrow-minrow
bias = minrow
print('----- Standard deviation -----')
print('c:',c)
#popt [2] save the new adjusted value of c
popt,pcov = curve_fit(gauss_function,np.arange(0,image_size[0]),sumf,p0=[a,b,c,bias])
print('Adjusted c:',popt[2])

plt.figure(5)
plt.plot(sumf,label='data')
plt.plot(gauss_function(np.arange(0,image_size[0]),*popt),'r-',label='fit')
plt.legend()
plt.title('Image row and gaussian fit')
plt.xlabel('Image row')
plt.ylabel('Row accumulated pixel value')
plt.show()
```

**ROI:** Una vez ajustada la función se selecciona una ROI que sea del doble del ajuste tanto hacia arriba como hacia abajo partiendo desde el máximo que se encontrará en el centro de la imagen y se muestra el resultado (Figura 5.13).

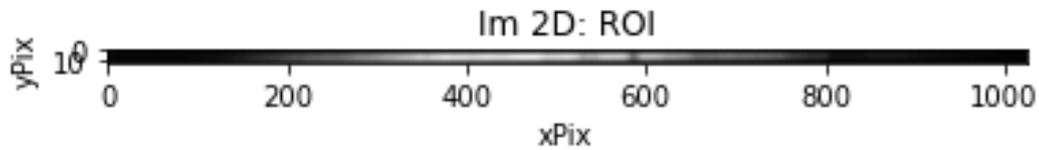

**Figura 5.13:** ROI.

Código 5.10: Selección de la ROI
```
#One time I know the width of the gaussian [popt[2]], I select the ROI we want
c_f = int(abs(2*popt[2])); #I choose the double of the width
ROI = image_data[int(localmaxima)-c_f:int(localmaxima)+c_f,:]
#Maximum +- double of the width

plt.figure(4)
plt.imshow(ROI,cmap='gray')
plt.xlabel('xPix')
plt.ylabel('yPix')
plt.title('Im 2D: ROI')
```

**Segunda aproximación a la gráfica del espectro:** Se vuelve a realizar el mismo paso que en la primea aproximación, pero en este caso se utiliza la ROI en lugar de la imagen completa. De esta forma se evitan todos los datos del resto de la imagen que no interesan y que pueden distorsionar el resultado.

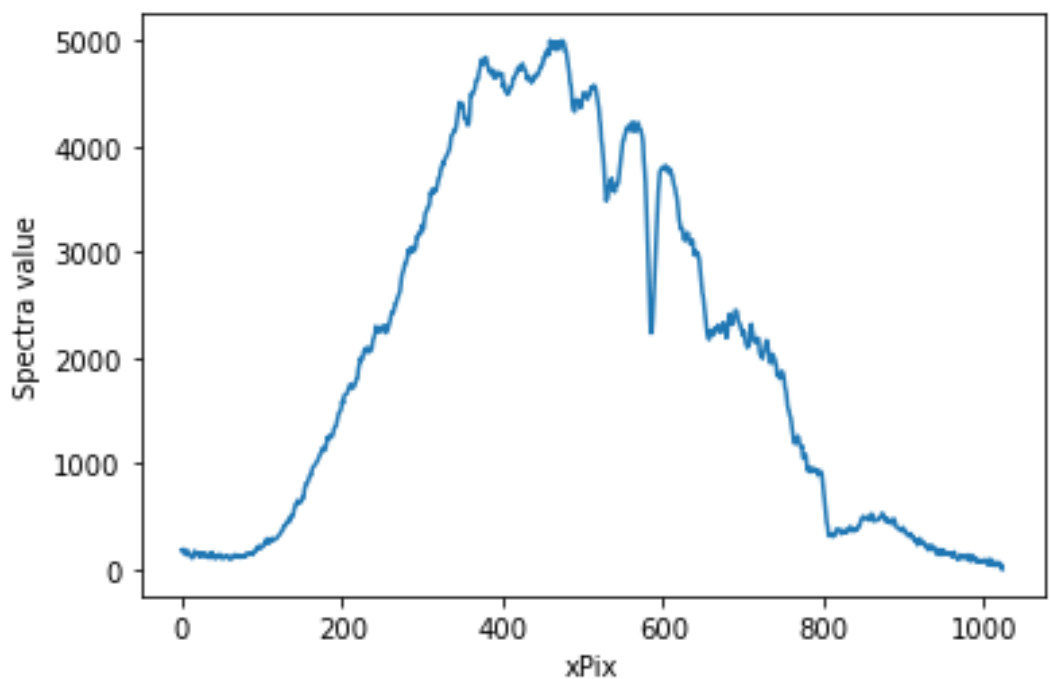

**Figura 5.14:** Espectro formado a partir de la ROI.

Código 5.11: Representación del espectro de la ROI

```
#To find the spectrum of this ROI
spectra = np.sum(ROI,axis=0)
plt.figure(5)
plt.plot(spectra) #I can see how the slope and absorption points are more evident here.
plt.xlabel('xPix')
plt.ylabel('Spectra value')
```

A diferencia de en la primera aproximación, en la (Figura 5.14) se pueden observar con mayor nitidez el continuo del espectro, así como sus líneas de absorción y emisión también se vuelven más evidentes.

**Elección de la configuración:** Tras obtener la gráfica del espectro respecto de la ROI solo falta expresar su eje X en longitud de onda (Angstrom). Esta es la finalidad de la lámpara de calibración *HgAr*.

Se disponen dos métodos distintos:

1. Introducir los datos extraídos de la tabla utilizada en una publicación anterior a este TFG y que caracteriza esta lampara (Figura 5.15) [13].

2. La opción más precisa es cargar el .txt del script *calibration.py* y utilizar sus datos.

| CCD x-coordinate | Fitted λ | Lab λ | note |
|---|---|---|---|
| 102.9 | 4046.7 | 4046.563 | HgI |
| 147.2 | 4358.0 | 4358.33 | HgI |
| 301.5 | 5462.8 | 5460.735 | HgI |
| 345.2 | 5778.8 | 5781.088 | HgId |
| 505.8 | 6963.7 | 6965.431 | ArI |
| 519.3 | 7066.0 | 7067.218 | ArI |
| 547.3 | 7279.7 | 7272.94 | Ar |
| 561.6 | 7389.2 | 7383.98 | Ar |
| 577.7 | 7513.3 | 7514.652 | Ar |
| 592.7 | 7629.1 | 7635.106 | Ar |
| 604.6 | 7721.3 | 7723.761 | ArI |
| 638.7 | 7986.4 | – | 2nd order HgI 4046 |
| 654.9 | 8112.4 | 8115.311 | ArI |
| 674.7 | 8266.4 | 8264.522 | ArI |
| 693.6 | 8413.6 | – | 2nd order HgI 4358 |
| 707.8 | 8524.4 | 8521.442 | ArI |
| 785.4 | 9123.5 | 9122.966 | ArI |
| 798.4 | 9223.3 | 9224.498 | ArI |
| 853.4 | 9649.2 | – | 2nd order Hg 5460 / refl. |
| 914.1 | 10139.8 | 10139.76 | HgI |
| 977.4 | 10695.4 | – | 2nd order Hg 5460 / refl. |

**Figura 5.15:** Tabla con la relación píxel-longitue de onda.

Código 5.12: Método de configuración

```python
#I can choose this method if the config_path is not available
if not config_path:

    #Pixels of Martin's table
    xHgAr = np.array([102,144,298,501,515,557,573,590,602,653,673,707,784,798,853])
    yHgAr = np.array([4046.563,4358.33,5460.735,6965.431,7067.218,7383.98,7514.652,7635.106,7723.761,8115.311,8264.522,8521.442,9122.96

#Or I can use the config_path to do the interpolation pixels-Amstrongs
else:

    xHgAr = np.loadtxt(config_path, max_rows=1)
    yHgAr = np.loadtxt(config_path, skiprows=1)
```

**Línea de interpolación:** Una vez se tienen los datos de los píxeles y sus correspon-dientes longitudes de onda se realiza la interpolación. Consiste en ver gráficamente que cada píxel tiene asociada una longitud de onda y que la relación es lineal (Figura 5.16).

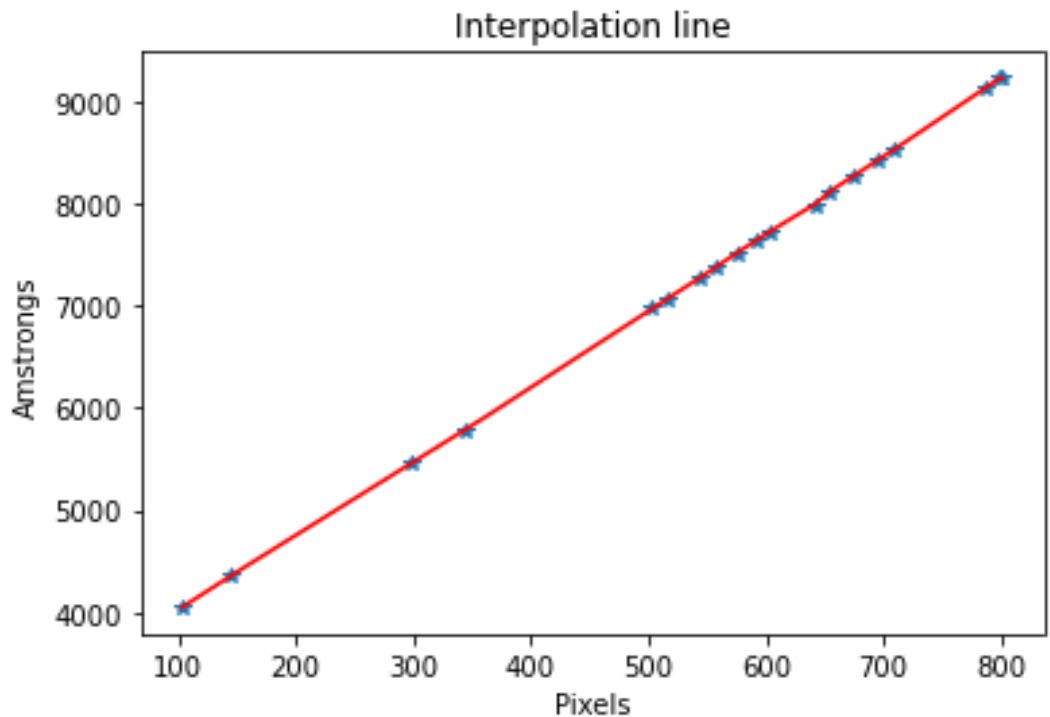

**Figura 5.16:** Linea de interpolación.

Código 5.13: Gráfica de la interpolación
```
1   #Show the interpolation
2   plt.figure(5)
3   x = np.linspace(np.min(xHgAr),np.max(xHgAr),1024)
4   #Define Y axe
5   y = np.interp(x,xHgAr,yHgAr)
6   #Plot points x respect y
7   plt.plot(xHgAr,yHgAr,'*')
8   #Plot line x respect y
9   plt.plot(x,y,'r')
10  plt.xlabel('Pixels')
11  plt.ylabel('Amstrongs')
12  plt.title('Interpolation line')
```

**Interpolación de cada píxel y espectro final:** Finalmente, se define una ecuación de primer grado con coeficientes a y b, siendo *xHgAr* el parámetro independiente donde los datos son medidos e *yHgAr* el parámetro dependiente.

Por último, se muestra la gráfica del espectro final con el eje X en *Angstroms* (Figura 5.17).

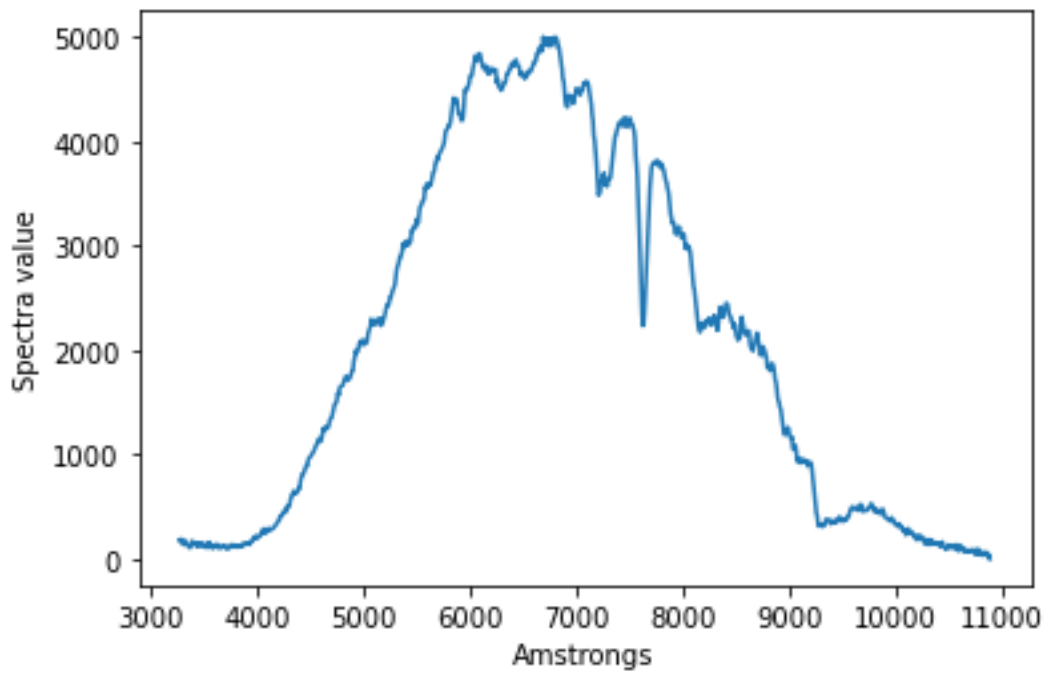

**Figura 5.17:** Espectro final en Angstroms.

Código 5.14: Interpolación de cada píxel y representación final del espectro

```
#I find the function to know every pixel's longwave
def func (x,a,b):
    return a*x+b
#Returns the coefficients of the function
popt,pcov = curve_fit(func,xHgAr,yHgAr)
a=(popt[0])
b=(popt[1])
#I call the function
lam=func(x,a,b)

#To find the spectrum of this ROI with x axe in amstrongs
plt.figure(6)
plt.plot(lam,spectra) #I can see how the slope and absorption points are more evident here.
plt.xlabel('Amstrongs')
plt.ylabel('Spectra value')
plt.savefig(graph_path, bbox_inches='tight')
```

### 5.3.2 calibration.py

**Definición de la función:** Al igual que en *spectrum.py*, se define la función en este caso para que creé un archivo *.txt*.

La función cuenta en esta ocasión con dos parámetros:

1. *fits path:* Indica la ruta donde se encuentra ubicada dentro del equipo la imagen

de la lámpara de calibración.

2. *config path:* Se encarga de crear fichero *.txt* donde se guardará la información de la lámpara de calibración.

Código 5.15: Definición de la función calibration
```
#I define the function calibration
def calibration(fits_path,config_path):
```

**Librerías:** Se definen las librerías que se van a utilizar a lo largo del programa.

Código 5.16: Librerías utilizadas
```
#Libraries I need to use
from astropy.io import fits
from scipy import ndimage
import numpy as np
from scipy.signal import find_peaks
import matplotlib.pyplot as plt
```

**Apertura del FITS y calibración:** Este paso es exactamente igual a su homologo en *spectrum.py* con la salvedad de que no es necesario guardar la dimensión del array y que no hay que utilizar el bias solo realizar el filtro de la mediana.

Código 5.17: FITS y calibración
```
#To working with FIT, first of all open the directory
fit_open = fits.open(fits_path)
#I work with the .fit as a list
hdu_list = fit_open
hdu_list.info()

image_data_ori = np.flip(hdu_list[0].data*1.0) #Turn data to float, in order to
#avoid overflow
#Median filter with a 2-pixel window, avoids hot spots
image_data = ndimage.median_filter(image_data_ori,(2))
```

**ROI y representación de la lámpara:** A continuación, se selecciona una ROI para reducir el tamaño de la imagen y se representa el espectro de la lámpara de calibración.

Seguidamente, se buscan los picos de la lámpara mediante la función *find peaks*, marcando una altura en el eje Y, que distinga los picos del continuo del espectro. Estos se representan con cruces azules (Figura 5.18).

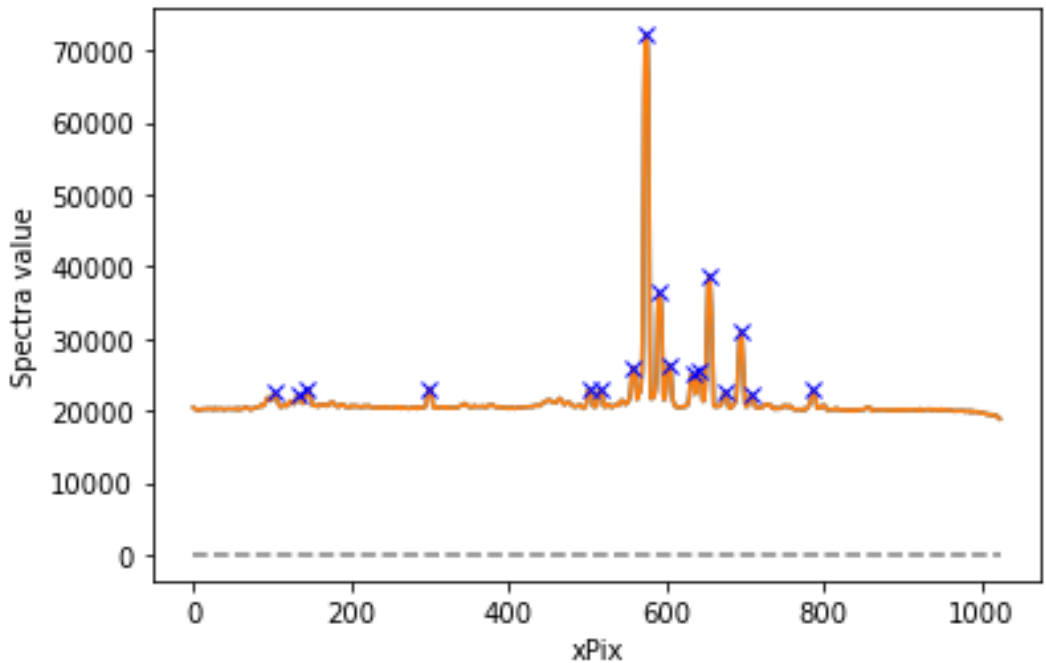

**Figura 5.18:** Espectro de la lampara HgAr con picos marcados.

Código 5.18: Representación de los picos de la lámpara

```
1   #Select a ROI to take lamps values
2   ROI = image_data[412:612,:]
3
4   #To find the spectrum of this ROI
5   spectra = np.sum(ROI,axis=0)
6   plt.plot(spectra)
7   plt.xlabel('xPix')
8   plt.ylabel('Spectra value')
9
10  #To show picks of our lamp FIT
11  peaks, _ = find_peaks(spectra, height=22000)
12  plt.plot(spectra)
13  plt.plot(peaks, spectra[peaks], "kx", color="blue")
14  plt.show()
```

**Guardado de los picos con su longitud de onda correspondiente:** Finalmente se guardan tanto los píxeles como las longitudes de onda en dos arrays distintos. Estos datos se van reordenando dentro de cada array mediante un bucle *for* y una serie de condicionales, de modo que cada píxel quedará en la misma posición del array que la longitud de onda que le corresponda.

Por último, se guardan en un archivo *.txt* tanto la posición de píxeles que forman máximos, como sus longitudes de onda en dos filas separadas.

Código 5.19: Correlación píxel longitud de onda y guardado

```python
#Save the values in 2 np.arrays
xHgAr = peaks
yHgAr = np.array([4046.563,4358.33,5460.735,5781.088,6965.431,7067.218,
7272.94,7383.98,7514.652,7635.106,7723.761,7986.4,8115.311,8264.522,8413.6,
8521.442,9122.966,9224.498,9649.2,10139.76,10695.4])

c=0 #counter

#I run the distance from 0 to the last saved peak
for i in range(0,len(xHgAr)):
    #Conditions for saving the data
    if (xHgAr[i] >= 100 and xHgAr[i] <= 106):
        xHgAr[c] = xHgAr[i]
        yHgAr[c] = 4046.563
        c+=1
    elif (xHgAr[i] >= 144 and xHgAr[i] <= 150):
        xHgAr[c] = xHgAr[i]
        yHgAr[c] = 4358.33
        c+=1
    elif (xHgAr[i] >= 298 and xHgAr[i] <= 305):
        xHgAr[c] = xHgAr[i]
        yHgAr[c] = 5460.735
        c+=1
    elif (xHgAr[i] >= 342 and xHgAr[i] <= 348):
        xHgAr[c] = xHgAr[i]
        yHgAr[c] = 5781.088
        c+=1
    elif (xHgAr[i] >= 502 and xHgAr[i] <= 509):
        xHgAr[c] = xHgAr[i]
        yHgAr[c] = 6965.431
        c+=1
    elif (xHgAr[i] >= 516 and xHgAr[i] <= 522):
        xHgAr[c] = xHgAr[i]
        yHgAr[c] = 7067.218
        c+=1
    elif (xHgAr[i] >= 544 and xHgAr[i] <= 550):
        xHgAr[c] = xHgAr[i]
        yHgAr[c] = 7272.94
        c+=1
    elif (xHgAr[i] >= 558 and xHgAr[i] <= 565):
        xHgAr[c] = xHgAr[i]
        yHgAr[c] = 7383.98
        c+=1
    elif (xHgAr[i] >= 574 and xHgAr[i] <= 581):
        xHgAr[c] = xHgAr[i]
        yHgAr[c] = 7514.652
        c+=1
    elif (xHgAr[i] >= 589 and xHgAr[i] <= 596):
        xHgAr[c] = xHgAr[i]
        yHgAr[c] = 7635.106
        c+=1
    elif (xHgAr[i] >= 601 and xHgAr[i] <= 608):
        xHgAr[c] = xHgAr[i]
        yHgAr[c] = 7723.761
        c+=1
    elif (xHgAr[i] >= 635 and xHgAr[i] <= 642):
        xHgAr[c] = xHgAr[i]
        yHgAr[c] = 7986.4
        c+=1
    elif (xHgAr[i] >= 651 and xHgAr[i] <= 658):
        xHgAr[c] = xHgAr[i]
        yHgAr[c] = 8115.311
        c+=1
```

```
64        elif (xHgAr[i] >= 671 and xHgAr[i] <= 678):
65            xHgAr[c] = xHgAr[i]
66            yHgAr[c] = 8264.522
67            c+=1
68        elif (xHgAr[i] >= 690 and xHgAr[i] <= 697):
69            xHgAr[c] = xHgAr[i]
70            yHgAr[c] = 8413.6
71            c+=1
72        elif (xHgAr[i] >= 704 and xHgAr[i] <= 711):
73            xHgAr[c] = xHgAr[i]
74            yHgAr[c] = 8521.442
75            c+=1
76        elif (xHgAr[i] >= 782 and xHgAr[i] <= 789):
77            xHgAr[c] = xHgAr[i]
78            yHgAr[c] = 9122.966
79            c+=1
80        elif (xHgAr[i] >= 795 and xHgAr[i] <= 801):
81            xHgAr[c] = xHgAr[i]
82            yHgAr[c] = 9224.498
83            c+=1
84        elif (xHgAr[i] >= 850 and xHgAr[i] <= 856):
85            xHgAr[c] = xHgAr[i]
86            yHgAr[c] = 9649.2
87            c+=1
88        elif (xHgAr[i] >= 911 and xHgAr[i] <= 917):
89            xHgAr[c] = xHgAr[i]
90            yHgAr[c] = 10139.76
91            c+=1
92        elif (xHgAr[i] >= 974 and xHgAr[i] <= 980):
93            xHgAr[c] = xHgAr[i]
94            yHgAr[c] = 10695.4
95            c+=1
96
97    #Save the peaks and their associated wavelengths into a .txt file
98    with open(config_path,"w") as f:
99        f.write("\n".join(" ".join(map(str, x)) for x in (xHgAr[0:c,],yHgAr[0:c,])))
```

# 6 Pruebas y resultados

En este capítulo se mostrará el procesamiento de varios FITS tomados con el telescopio. La finalidad es demostrar que el software implementado es robusto y funciona sin fallos aparentes con distintos tipos de imágenes espectrales.

Para ello se han llevado a cabo diversas pruebas con imágenes de diversa índole. En este punto se mostrarán los resultados más interesantes.

## 6.1 Linea de absorción falsas

Uno de los fenómenos que se ha manifestado a lo largo de todas la pruebas es el una falsa línea de emisión producida por la propia mecánica del telescopio.

Como se explica en el apartado 3.2.1, es necesario el uso de flats para la calibración de las imágenes. En este caso, se generó un flatmaster (Figura 6.1) mediante código, pero no se ha implementado puesto que desvirtuaba el espectro final.

Como puede comprobarse, hay una línea negra que cubre toda la columna de los píxeles 575-580. Esto produce que cuando se toman las imágenes espectrales esa línea aparece generando una falsa línea de absorción (Figura 6.2).

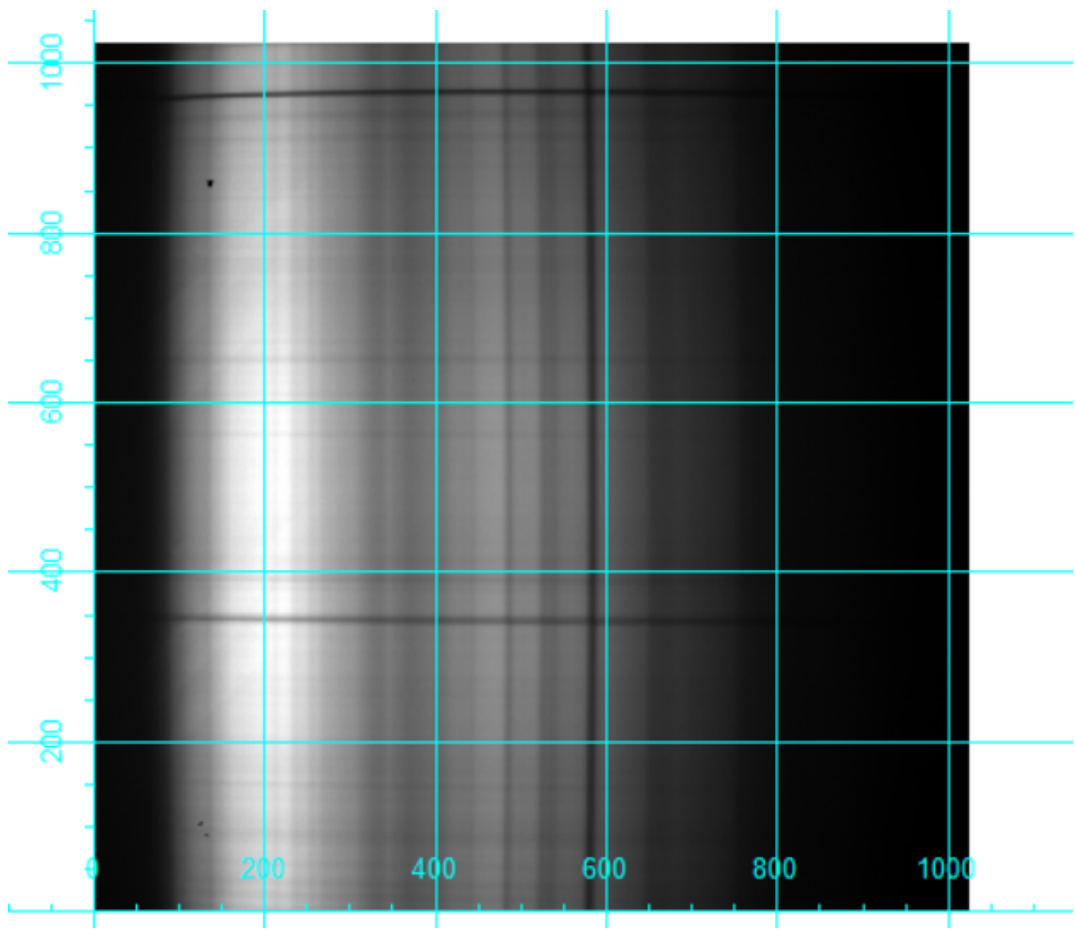

**Figura 6.1:** Flatmaster generado.

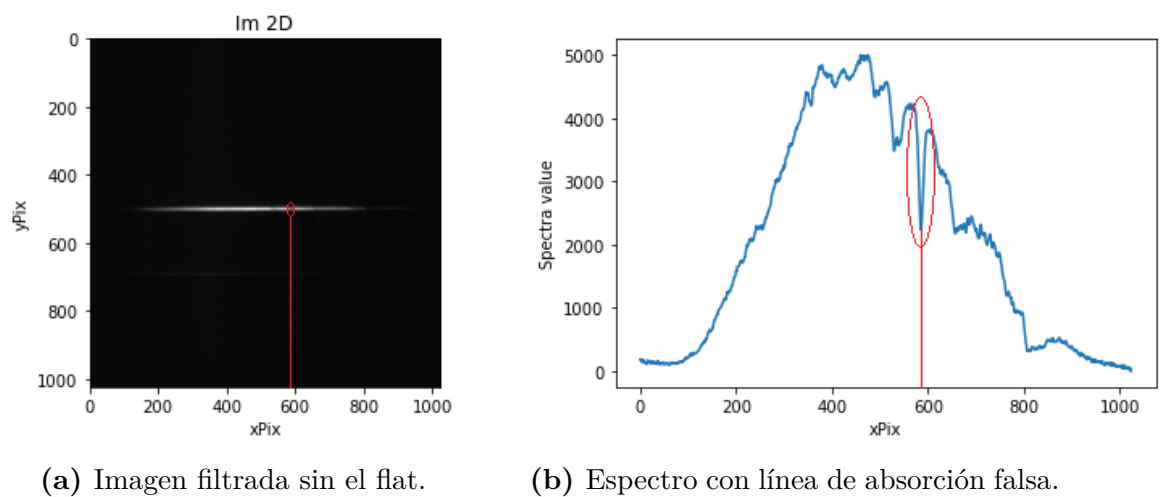

(a) Imagen filtrada sin el flat.   (b) Espectro con línea de absorción falsa.

**Figura 6.2:** Linea falsa generada por la óptica del telescopio.

## 6.2 Espectro de interés

En algunas ocasiones puede darse el fenómeno de que dos estrellas se encuentren alineadas verticalmente en el centro de la imagen, como se muestra en la (Figura 6.3).

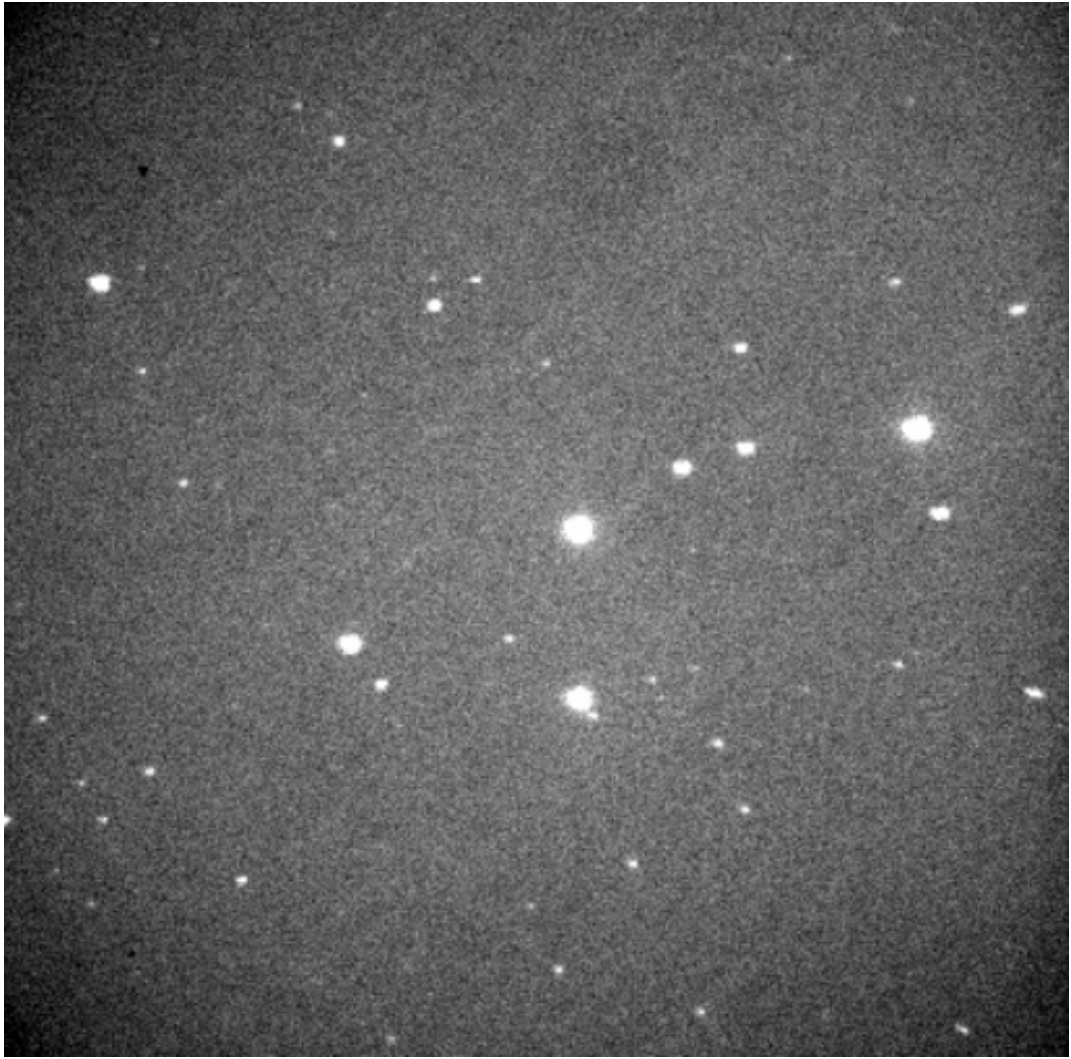

**Figura 6.3:** Ejemplo de dos estrellas alineadas verticalmente tomado con BOOTES 2.

Esto provoca que al colocar la rendija, la luz de ambas quede capturada en la imagen espectral como se muestra en la (Figura 6.4).

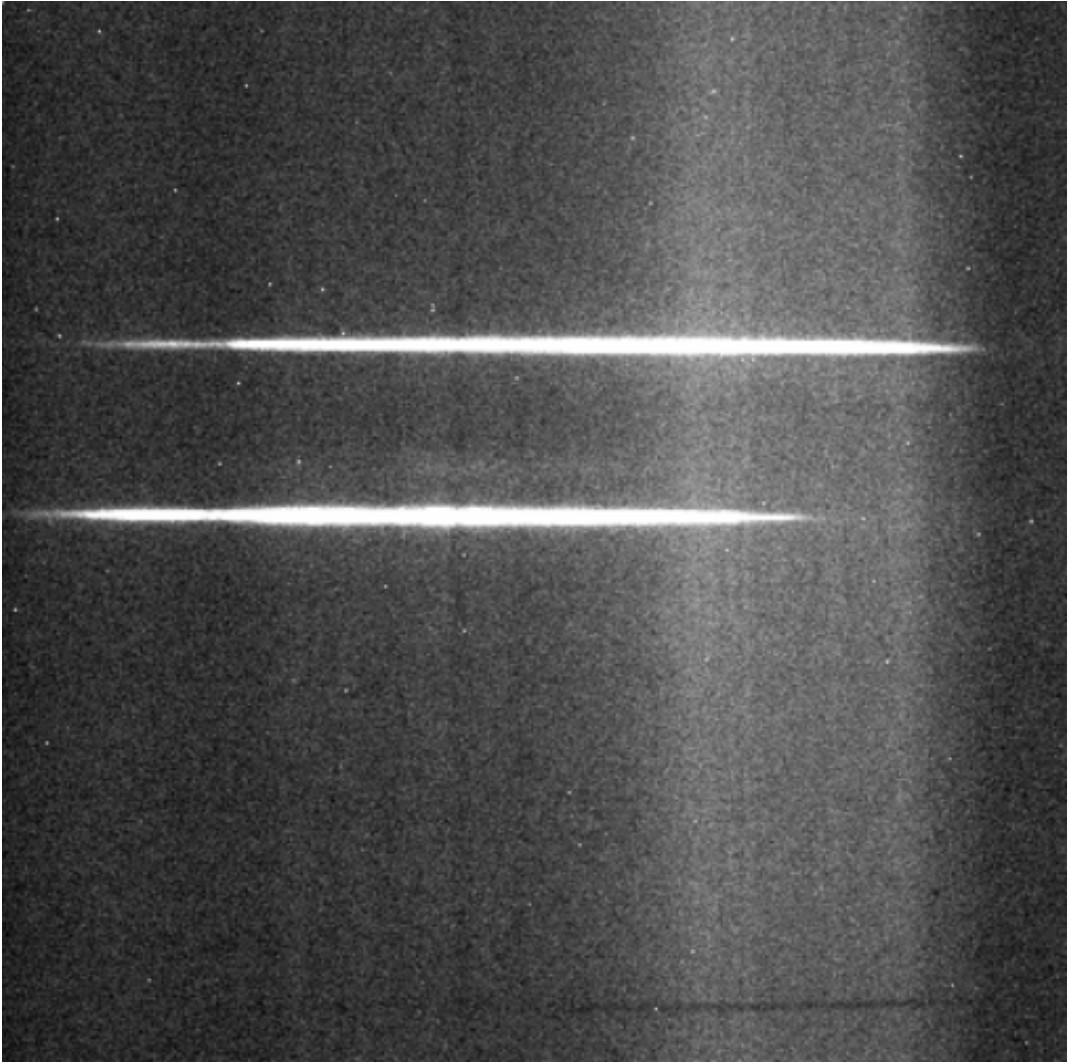

**Figura 6.4:** Imagen espectral de dos estrellas tomada con BOOTES 2.

En la (Figura 6.5) vemos como esto no supone un problema, pues el software es capaz de filtrar de forma que solo se queda con el espectro alineado en el centro de la imagen, al tener este el pico máximo.

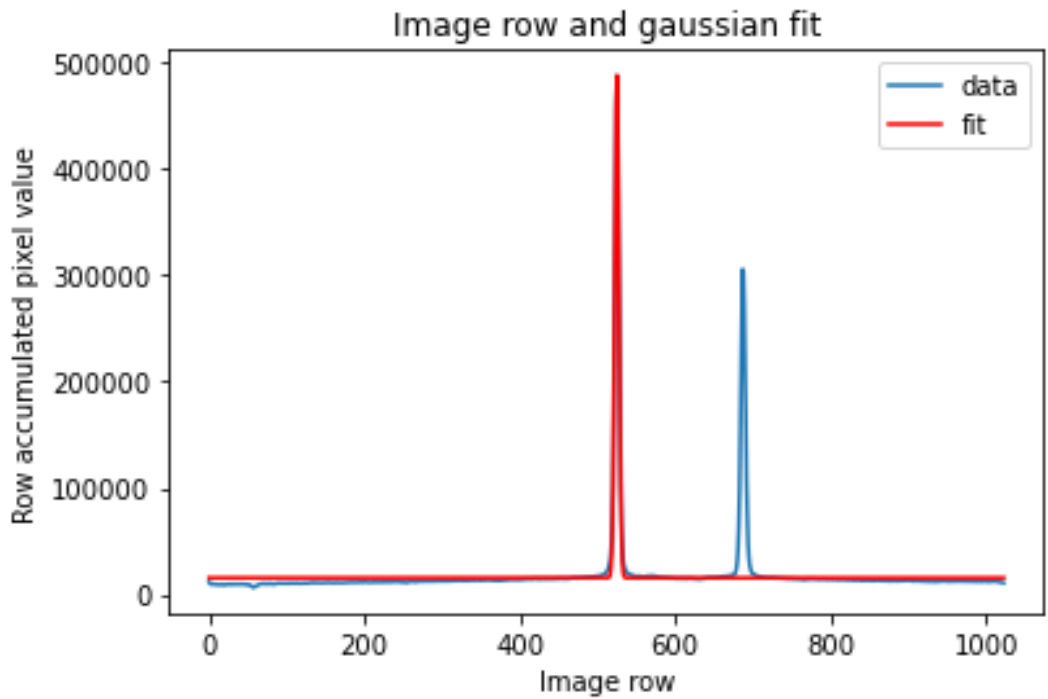

**Figura 6.5:** Ajuste gaussiano del espectro de interés.

## 6.3 Fenómenos estudiados

### 6.3.1 Novas

Alguno de los fenómenos estelares más espectaculares y a su vez más difíciles de observar son las novas.[1] A continuación, se mostrará el resultado de la observación de algunas de ellas.

**Nova Cas 2021:** Conocida formalmente como V1405 Cas por la Unión astronómica internacional (UAI) fue descubierta el 18 de marzo de 2021, cuando brillaba a una magnitud de +9,6. El espectro capturado con COLORES (Figura 6.6), se ha comparado con uno guardado en la base de datos de la British Astronomical Association (BAA) (Figura 6.7).

---

[1] Explosión termonuclear causada por un desequilibrio entre la gravedad de una estrella y su combustible nuclear.

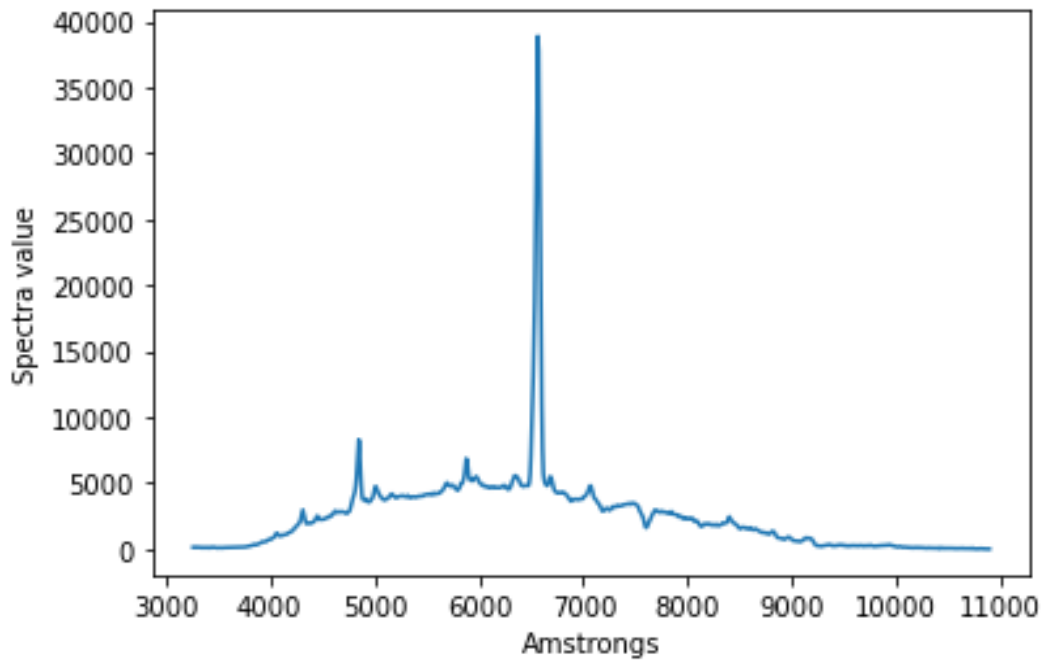

**Figura 6.6:** Espectro de V1405 Cas capturado con BOOTES 2.

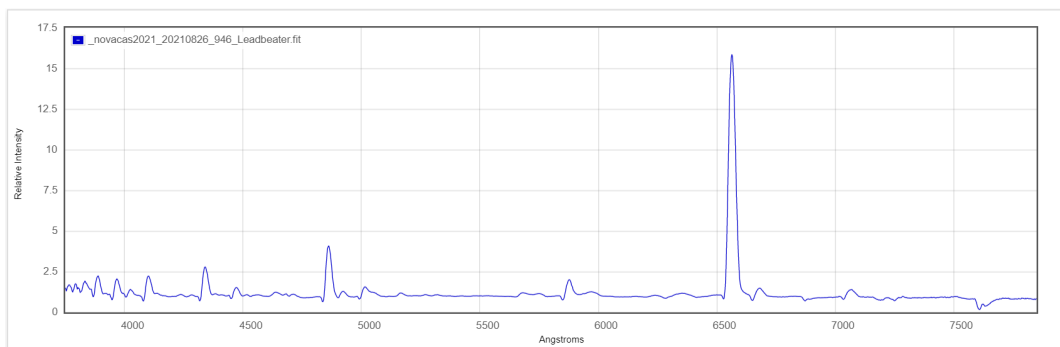

**Figura 6.7:** Espectro de V1405 Cas obtenido por el BAA.

Como se puede observar, son bastante similares aunque salvando las distancias ya que COLORES se trata de un espectrógrafo de baja resolución. Aún así, es lo suficientemente preciso para ver la gran línea de emisión en los 6560 Angstroms correspondiente al hidrógeno en la serie de Balmer, más concretamente a H-alfa. También se encuentran representadas las demás, pero esta es la más evidente.

**Nova Vul 2021:** Conocida formalmente como V606 Vul por la UAI fue descubierta el 16 de julio de 2021, cuando brillaba a una magnitud de +12. El espectro capturado con COLORES (Figura 6.8), se ha comparado con uno guardado en la base de datos de la BAA (Figura 6.9).

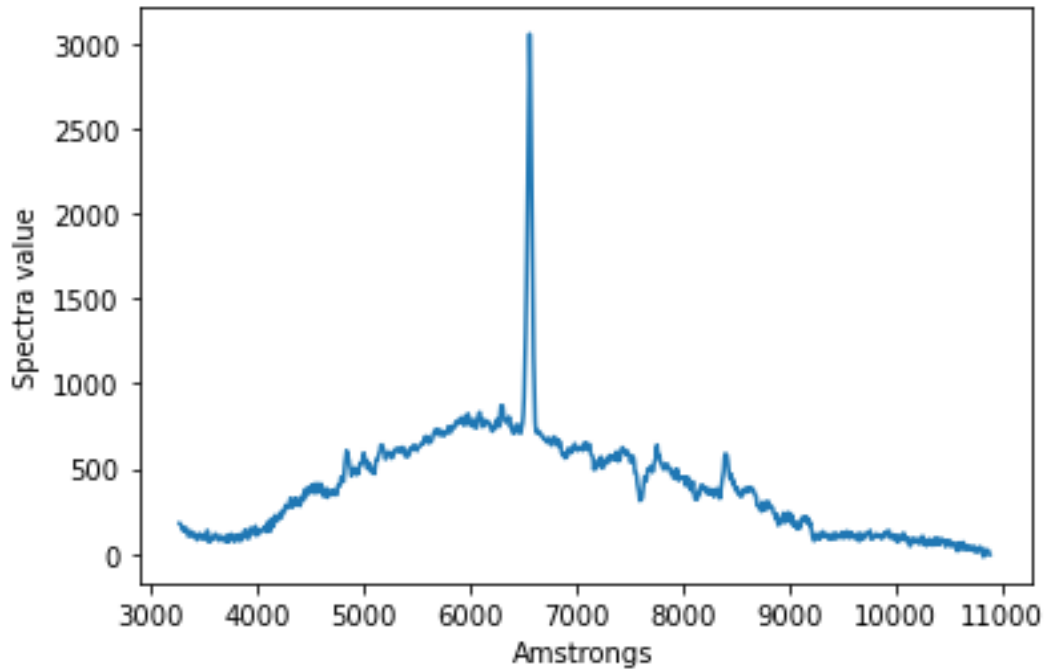

**Figura 6.8:** Espectro de V606 Vul capturado con BOOTES 2.

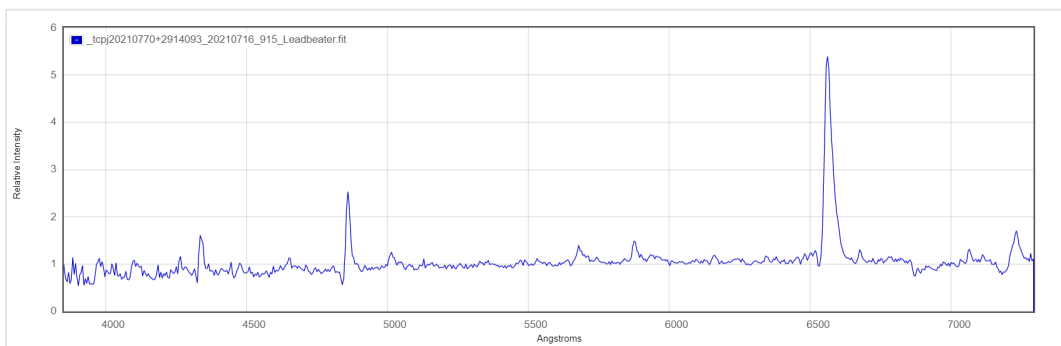

**Figura 6.9:** Espectro de V606 Vul obtenido por el BAA.

Al igual que en el caso anterior, se vuelve a demostrar que el script consigue procesar una gráfica muy cercana a la real donde quedan patentes las líneas de emisión del hidrógeno.

**Novas enanas**

Las novas enanas pertenecen al grupo de estrellas variables. Son unos sistemas muy curiosos pues pueden variar su magnitud de forma repentina. El aumento hasta hasta su pico máximo tiene lugar en menos de un día, mientras que la disminución hasta la inactividad sucede durante semanas.[14]

**SS Cygni**: Conocida formalmente como GCRV 13641 por la UAI, pertenece a la constelación de Cygnus a 330 años luz del Sistema Solar. Su magnitud aparente oscila entre un mínimo de +12,2 y un máximo de +8,3. Se trata de una de las estrellas variables más famosas y por lo tanto más observadas del firmamento. El espectro capturado con COLORES (Figura 6.10), se ha comparado con uno guardado en la base de datos de la BAA (Figura 6.11)

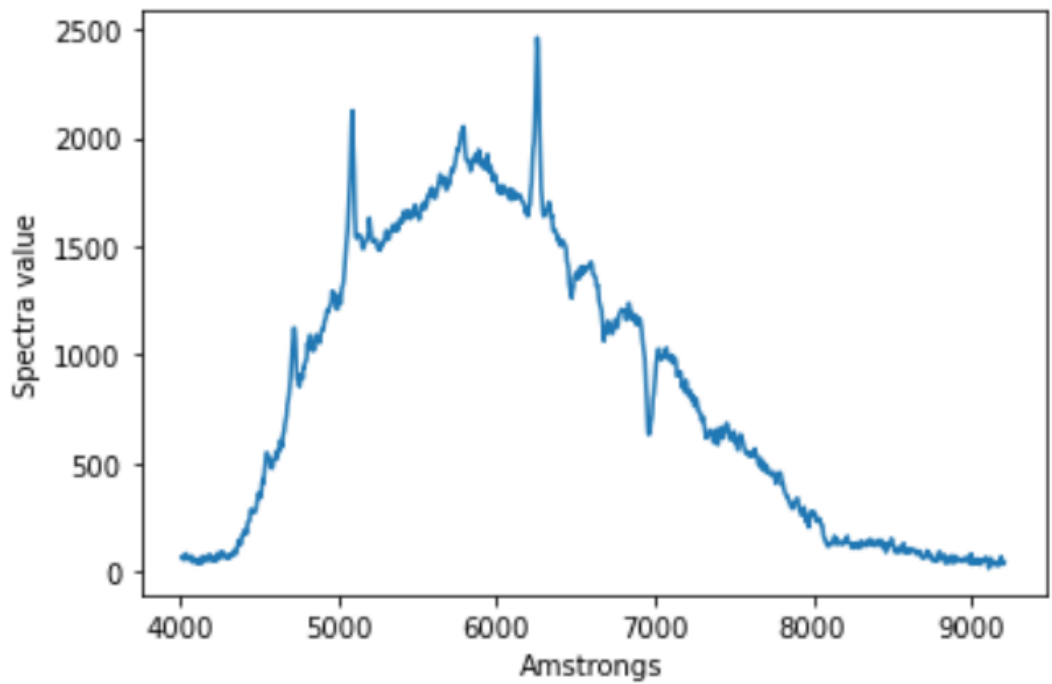

**Figura 6.10:** Espectro de SS Cygni capturado con BOOTES 2.

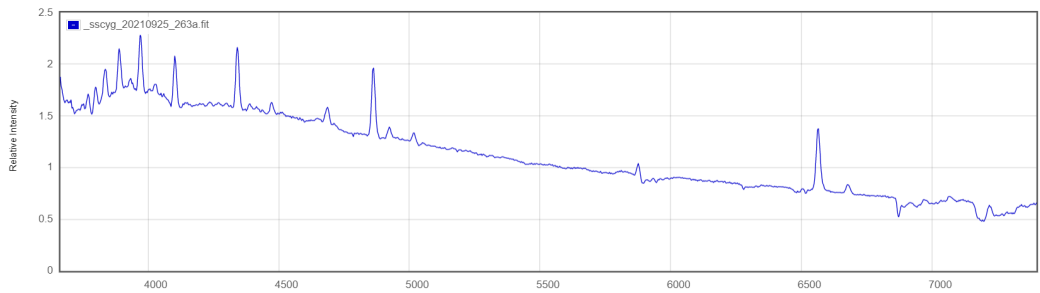

**Figura 6.11:** Espectro de SS Cygni obtenido por el BAA.

Se puede comprobar como sobre los se repiten las lineas de emisión a los 5000 y 6560 Angstroms correspondiente al hidrógeno en la serie de Balmer, más concretamente a H-alfa y al H-beta. Al igual que en casos anteriores el resto son menos visibles.

### 6.3.2 Estrella simbiótica

Se trata de un sistema estelar binario normalmente formado por una enana blanca y una gigante roja. Las explosiones (novas) se producen cuando la enana blanca acumula o extrae demasiada masa de su compañera gigante roja. El exceso de masa desencadena una explosión termonuclear en una capa de hidrógeno en la superficie de la enana blanca [15]. Son de gran interés científico pues otorgan una gran cantidad de información sobre la evolución estelar.

**RS Oph:** Conocida formalmente como RS Ophiuchi por la UAI. RS Oph forma parte de un sistema estelar de novas recurrentes. El espectro capturado con COLORES (Figura 6.12), se ha comparado con uno guardado en la base de datos de la BAA (Figura 6.13).

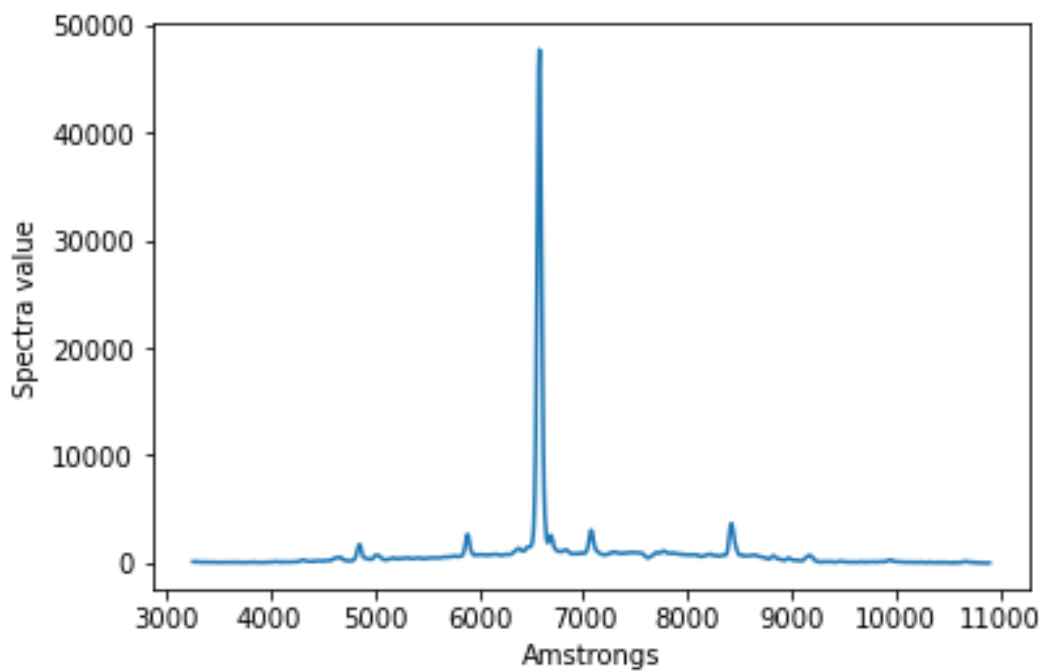

**Figura 6.12:** Espectro de RS Ophiuchi capturado con BOOTES 2.

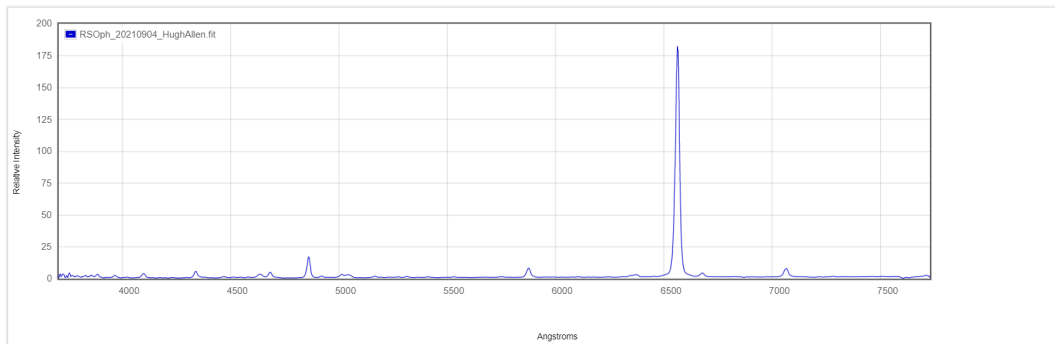

**Figura 6.13:** Espectro de RS Ophiuchi obtenido por el BAA.

De igual modo a como ocurre con las novas anteriores también son muy evidentes en este caso las líneas de emisión pertenecientes a las series de Balmer del hidrógeno.

**V1016 Cyg:** Pertenece a la constelación del cisne a unos 6000 años luz de distancia del Sistema Solar. En esta ocasión se muestra la imagen FITS descargada de la web del telescopio para denotar la exactitud del espectro (Figura 6.14). El espectro capturado con COLORES (Figura 6.15), se ha comparado con uno guardado en la base de datos de la BAA (Figura 6.16).

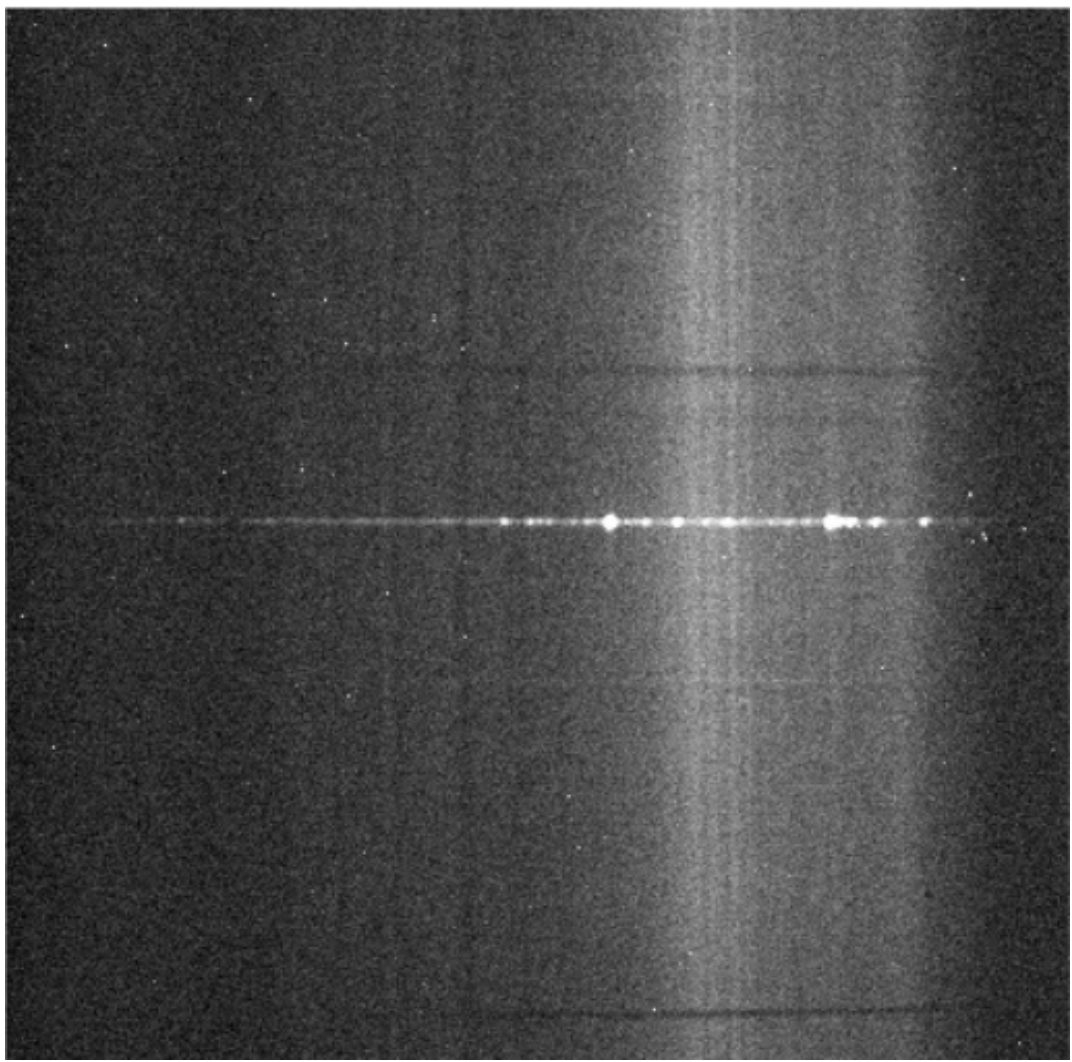

**Figura 6.14:** FIT de V1016 Cyg capturado con BOOTES 2.

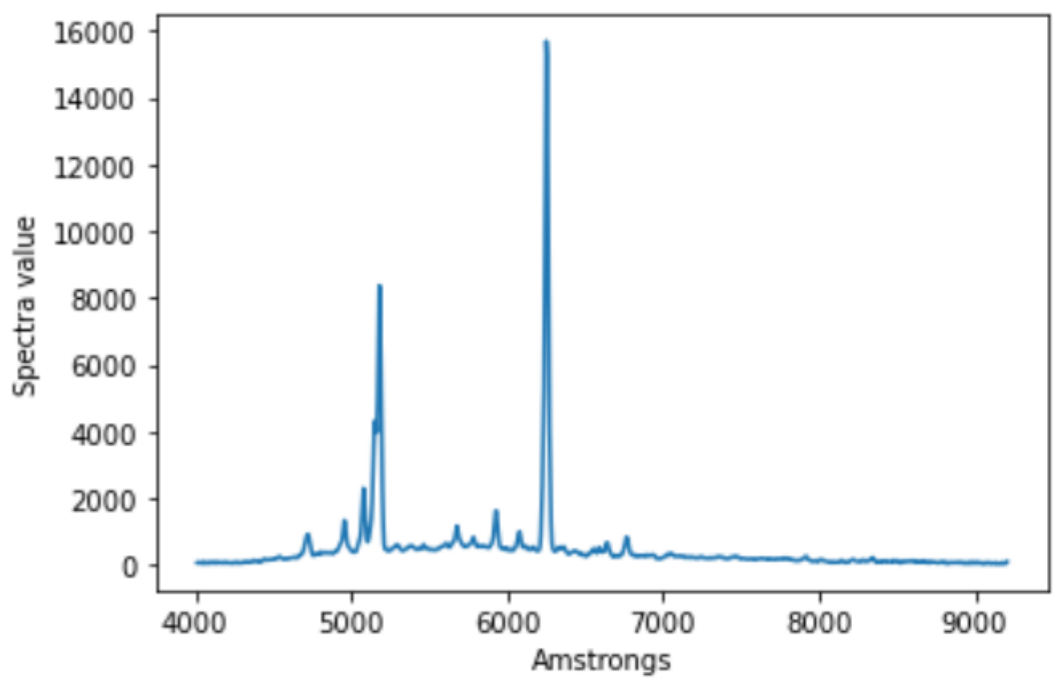

**Figura 6.15:** Espectro de V1016 Cyg capturado con BOOTES 2.

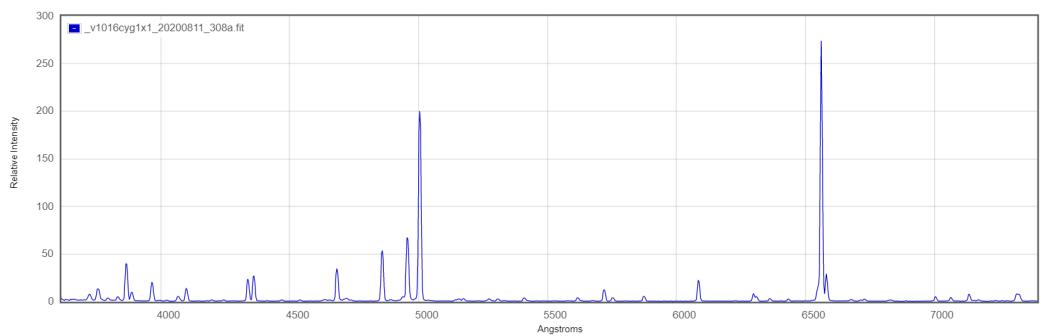

**Figura 6.16:** Espectro de V1016 Cyg obtenido por el BAA.

En este caso se aprecia con gran exactitud las lineas de emisión del Hidrógeno alpha y beta.

**StHA 190:** El espectro capturado con COLORES (Figura 6.17), se ha comparado con uno guardado en la base de datos de la BAA (Figura 6.18).

**Figura 6.17:** Espectro de StHA 190 capturado con BOOTES 2.

**Figura 6.18:** Espectro de StHA 190 obtenido por el BAA.

De igual modo a como ocurre en el caso anterior también son muy evidentes en este caso las líneas de emisión pertenecientes a las series de Balmer del hidrógeno.

### 6.3.3 Estrellas

**Estrellas Be**

Son estrellas de tipo espectral Be cuyo espectro tiene, o tuvo en algún momento, una o más líneas de emisión de la serie de Balmer. Aunque el espectro de tipo Be se produce con mayor intensidad en las estrellas de clase B, también se detecta en estrellas de cascarón O y A.

**EE Cep:** Es una de las estrellas tipo B de período más largo, con un período de unos 2050 d. EE Cep está formada por la estrella azul B5ne y una compañera invisible. El espectro capturado con COLORES (Figura 6.19), se ha comparado con uno guardado en la base de datos de la BAA (Figura 6.20).

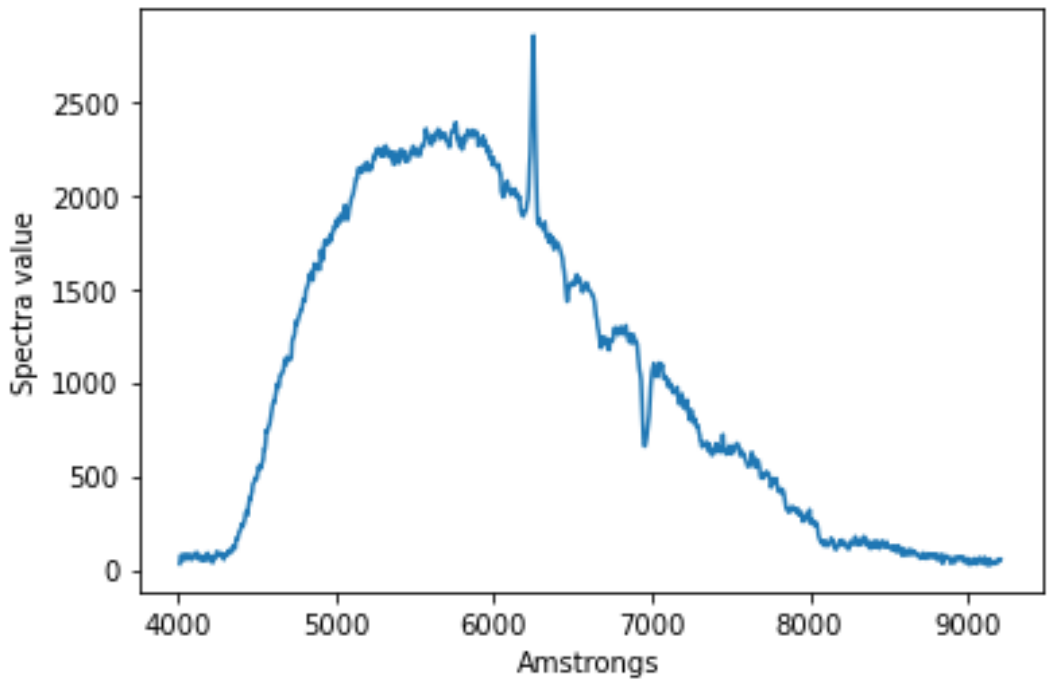

**Figura 6.19:** Espectro de EE Cep capturado con BOOTES 2.

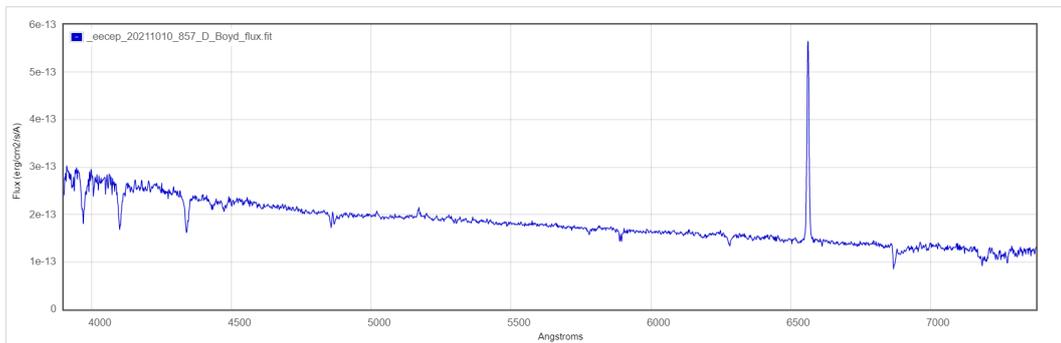

**Figura 6.20:** Espectro de EE Cep obtenido por el BAA.

Se comprueba una clara linea de emisión en los 6500 Angstroms perteneciente al

H-alpha.

**Estrellas de carbono**

Este tipo de estrellas suelen caracterizarse por tratarse de una gigante roja luminosa, cuya atmósfera contiene más carbono que oxígeno. Estos elementos forman en la capa superior de la estrella monóxido de carbono. Este compuesto químico consume el oxígeno presente en la atmósfera, dejando que el carbono tinte la capa externa de la estrella de color rojo rubí.[16]

**OR Cep:** Se encuentra situada a unos 45 años luz de distancia en la constelación de Cefeo. El espectro capturado con COLORES (Figura 6.21), se ha comparado con uno guardado en la base de datos de la BAA (Figura 6.22).

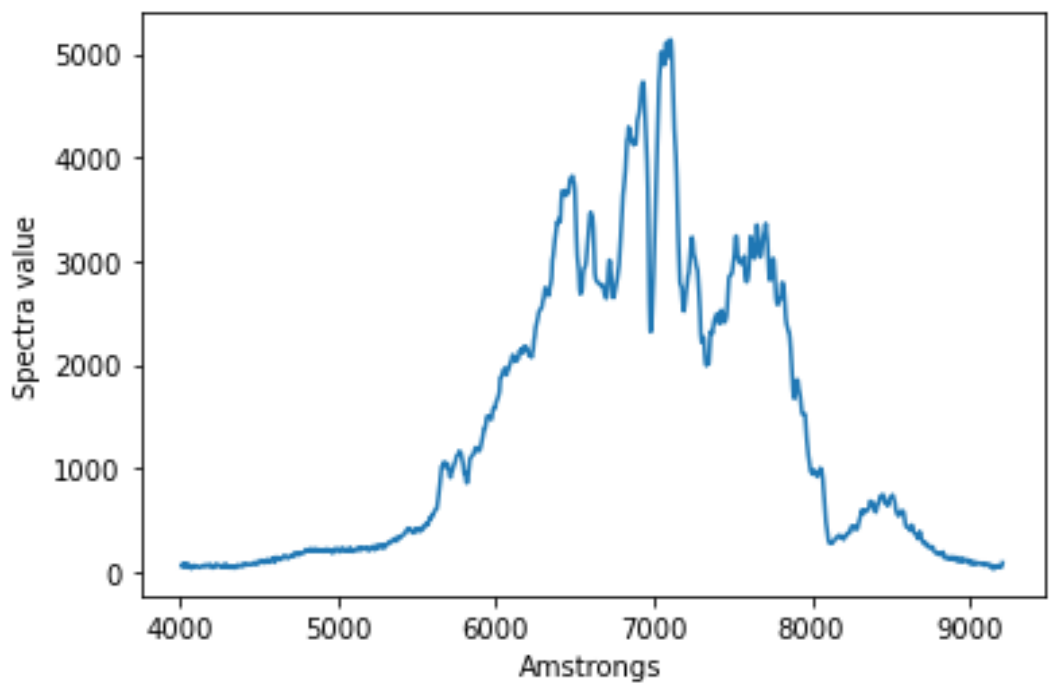

**Figura 6.21:** Espectro de OR Cep capturado con BOOTES 2.

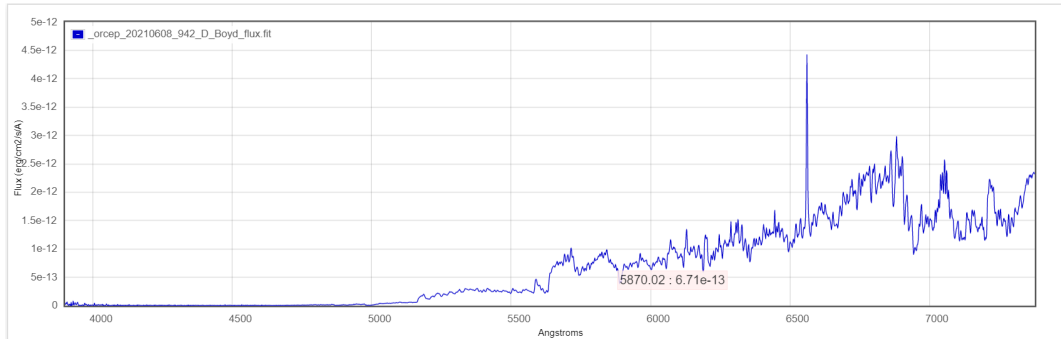

**Figura 6.22:** Espectro de OR Cep obtenido por el BAA.

Se comprueba una clara linea de emisión en los 6500 Angstroms perteneciente al H-alpha y posteriormente una gran línea de absorción cercana a los 7000 Angstroms probablemente perteneciente a B(O2).

**Otras estrellas**

**SAO 11483:** El espectro capturado con COLORES (Figura 6.23) muestra distintas líneas de absorción y emisión.

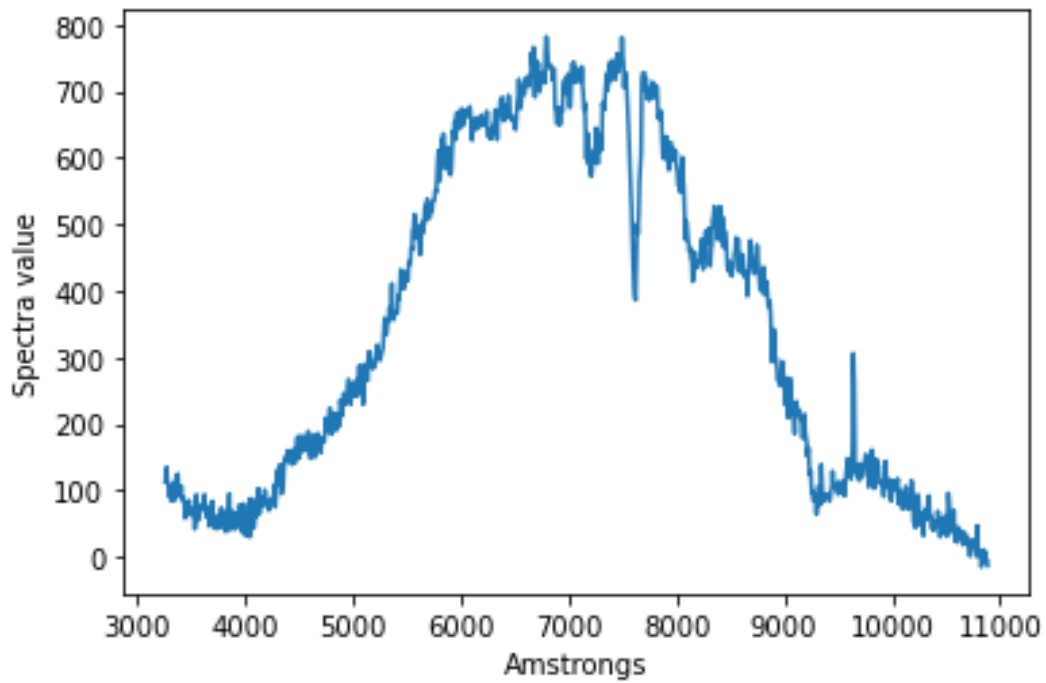

**Figura 6.23:** Espectro de SAO 11483 capturado con BOOTES 2.

**SAO 22966:** El espectro capturado con COLORES (Figura 6.24) muestra distintas líneas de absorción y emisión.

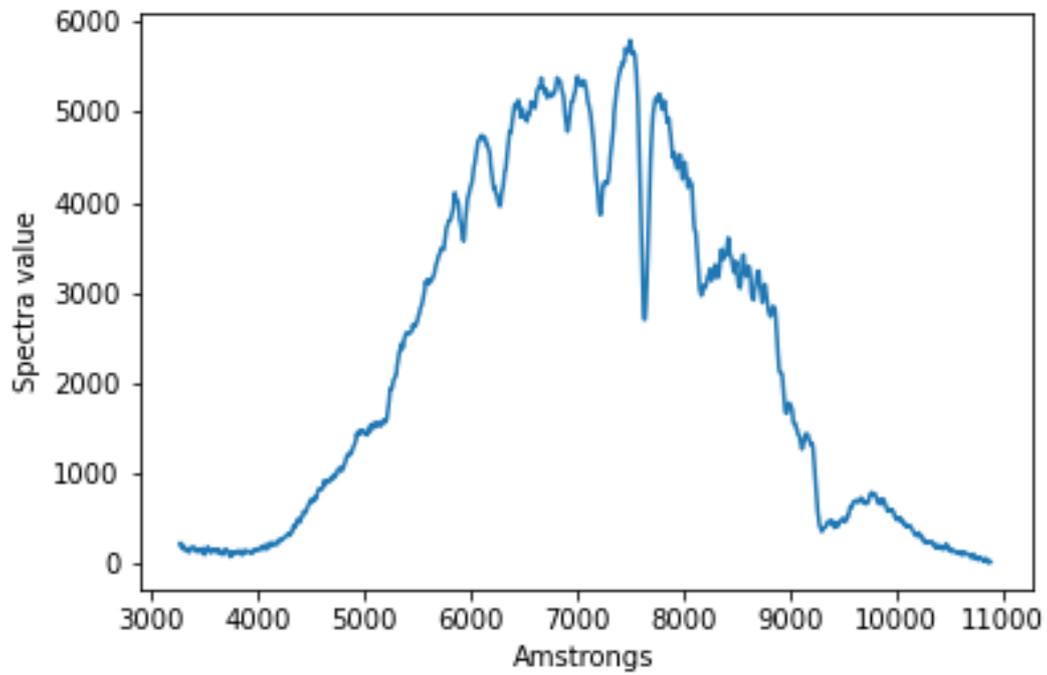

**Figura 6.24:** Espectro de SAO 22966 capturado con BOOTES 2.

**SAO 46525:** Pertenece a la constelación de Hércules. El espectro capturado con COLORES (Figura 6.25) muestra distintas líneas de absorción y emisión.

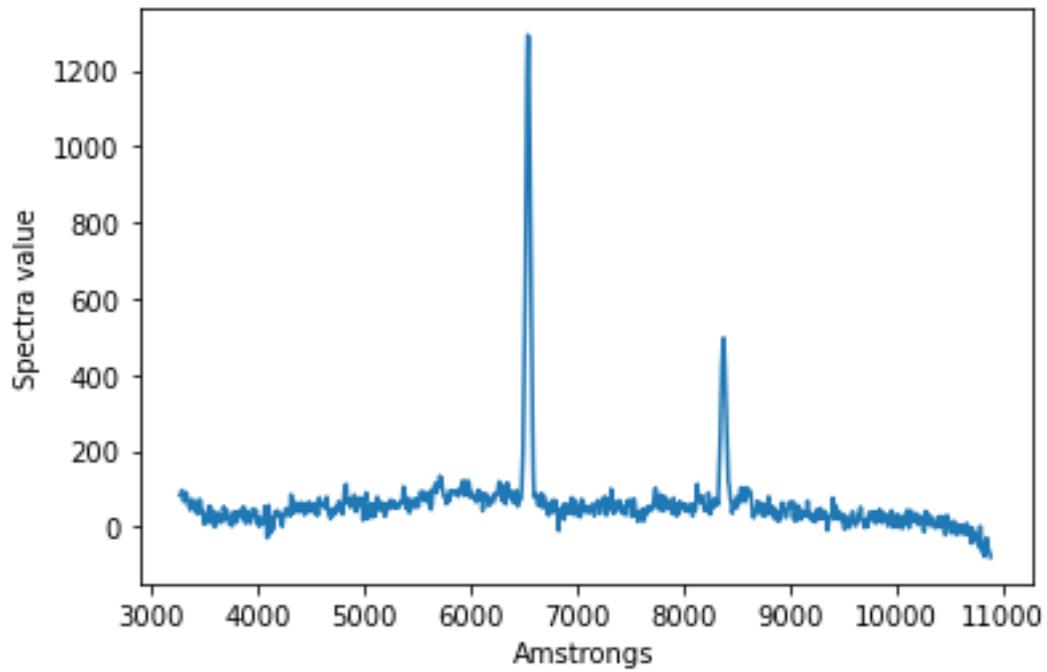

**Figura 6.25:** Espectro de SAO 46525 capturado con BOOTES 2.

Se constatan dos claras lineas de emisión. Una de ellas perteneciente a H- (alfa) y la otra no está identificada.

# 7 Conclusiones

El objetivo con el que se desarrolla este TFG es la puesta en marcha del espectrógrafo COLORES, la comprobación de su correcto funcionamiento y su implementación en la Web del telescopio mediante el software desarrollado. Aunque el software se podría depurar aún más mediante técnicas más sofisticadas, cumple con su propósito y permite comprobar su correcto funcionamiento, ofreciendo así en la estación BOOTES 2 una herramienta de enorme valor para la investigación.

Con el fin de lograr esta meta se han utilizado todos los conocimientos adquiridos durante la realización del grado, tanto en visión por computador como en adquisición de imágenes. Además, se ha aprendido durante el desarrollo del proyecto sobre el lenguaje de programación Python utilizado en el software. Por último, ha habido un trabajo de investigación y aprendizaje sobre el funcionamiento de telescopios, astronomía, etc...

Es por todos estos motivos que este proyecto ha supuesto una primera aproximación profesional a la ingeniería, poniendo sobre la mesa todos los conocimientos adquiridos tanto en astronomía como en instrumentación y programación. Logrando de esta manera una satisfacción personal a la altura de este reto.

## 7.1 Trabajos Futuros

Uno de los problemas que han surgido durante la realización de este TFG ha sido el de la calibración de las lámparas que sirven como referencia a la hora de caracterizar el espectro. Como futura implementación, se podría estudiar la verdadera naturaleza de las lámparas integradas en el telescopio y de este modo obtener de una forma más precisa el espectro de la imagen estudiada.

Otro aspecto a tener en cuenta es el del ruido presente en las imágenes tomadas por el telescopio. Se encontró durante las observaciones, que al colocar las rendijas se genera un ruido que interfiere en la imagen que se quiere estudiar. Esto se puede deber a un desgaste del prisma, puesto que con todas las rendijas en mayor o menor grado encontramos este flat. Es por eso, que otra implementación futura puede ser la de colocar un prisma nuevo que opere en perfecto estado y no distorsione los resultados, además de una puesta a punto del resto de subsistemas del espectrógrafo.

Por último, se podría añadir al código una rutina que localizara las distintas lineas de emisión y absorción del espectro estudiado. Mediante ellas y la consulta en una base de datos se podría obtener a que elemento pertenece cada línea de manera automatizada.

# Bibliografía

# Lista de Acrónimos y Abreviaturas

| | |
|---|---|
| **ASCII** | American Standard Code for Information Interchange. |
| **BAA** | British Astronomical Association. |
| **BOOTES** | Burst Observer and Optical Transient Exploring System. |
| **CCD** | Dispositivo de carga acoplada. |
| **COLORES** | Compact Low Resolution Spectrograp. |
| **DEC** | Declinación. |
| **FITS** | Flexible Image Transport System. |
| **FOSC** | Faint Object Spectrograph and Camera. |
| **GRB** | Gamma-Ray Burst. |
| **IAA** | Instituto de Astrofísica de Andalucía. |
| **RA** | Ascensión Recta. |
| **ROI** | Region of interest. |
| **TELMA** | Telescopio de Málaga. |
| **TFG** | Trabajo Final de Grado. |
| **UAI** | Unión astronómica internacional. |
| **UTC** | Tiempo Universal Coordinado. |

# Índice de figuras

# Índice de Códigos